\documentclass[acmtog, authorversion]{acmart}

\usepackage{graphicx}
\usepackage{enumitem}
\usepackage{calc}
\AtBeginDocument{%
  \providecommand\BibTeX{{%
    \normalfont B\kern-0.5em{\scshape i\kern-0.25em b}\kern-0.8em\TeX}}}

\begin{document}

% \title[Introducing Playful Human-Robot Collaboration as a Strategy]{Encouraging Bystander Assistance for Urban Robots: Introducing Playful Human-Robot Collaboration as a Strategy}

\title[Introducing Playful Robot Help-Seeking as a Strategy]{Encouraging Bystander Assistance for Urban Robots: Introducing Playful Robot Help-Seeking as a Strategy}

%Exploring Playful Engagements to Encourage Bystander Assistance for Urban Robots

% Enticing Bystanders to Assist Urban Robots: Play as a Strategy  
% Enticing Bystanders to Assist Urban Robots: Playful Engagement as a Strategy  
% Enticing Bystanders to Assist Urban Robots: Introducing Playful Engagement as a Strategy  
% Enticing Bystanders to Assist Urban Robots: Introducing Playful Human-Robot Collaboration as a Strategy  
%Encouraging Bystander Assistance for Urban Robots: Introducing Playful Human-Robot Collaboration as a Strategy  

% Investigating Design Strategies to Encourage Bystander Assistance for Urban Robots

% Introducing 

\author{Xinyan Yu}
\email{xinyan.yu@sydney.edu.au}
\orcid{0000-0001-8299-3381}
\affiliation{Design Lab,
  \institution{The University of Sydney}
  \city{Sydney}
  \state{NSW}
  \country{Australia}
}

\author{Marius Hoggenmüller}
\email{marius.hoggenmuller@sydney.edu.au}
\orcid{0000-0002-8893-5729}
\affiliation{Design Lab,
  \institution{The University of Sydney}
  \city{Sydney}
  \state{NSW}
  \country{Australia}
}

\author{Martin Tomitsch}
\email{Martin.Tomitsch@uts.edu.au}
\orcid{0000-0003-1998-2975}
\affiliation{Transdisciplinary School,
  \institution{University of Technology Sydney}
  \city{Sydney}
  \state{NSW}
  \country{Australia}
}


\renewcommand{\shortauthors}{Yu, et al.}

\begin{abstract}
Robots in urban environments will inevitably encounter situations beyond their capabilities (e.g., delivery robots unable to press traffic light buttons), necessitating bystander assistance. These spontaneous collaborations possess challenges distinct from traditional human-robot collaboration, requiring design investigation and tailored interaction strategies. This study investigates playful help-seeking as a strategy to encourage such bystander assistance. We compared our designed playful help-seeking concepts against two existing robot help-seeking strategies: verbal speech and emotional expression. To assess these strategies and their impact on bystanders' experience and attitudes towards urban robots, we conducted a virtual reality evaluation study with 24 participants. Playful help-seeking enhanced people's willingness to help robots, a tendency more pronounced in scenarios requiring greater physical effort. Verbal help-seeking was perceived less polite, raising stronger discomfort assessments. Emotional expression help-seeking elicited empathy while leading to lower cognitive trust. The triangulation of quantitative and qualitative results highlights considerations for robot help-seeking from bystanders.  
%The triangulation of quantitative and qualitative results further highlights design considerations for facilitating such casual collaboration.  

%the interplay of factors shaping the casual collaboration between bystanders and robots in urban spaces.

% 1. context/state of the art: What is the current state of the field? How are you positioning the research in relation to what exists?

% 2. issue/problem: What is a key issue? or a problem originating either from practice or previous research? 1 and 2 help to frame the research for the reader.

% 3. narrow in to the gap -> research question: Narrow down the focus and state as a research question.

% 4. your approach to the research: What methodology and methods did you apply? for what purpose?

% 5. results/key findings: Summarise the key findings and/or outcomes from your study. Be specific in describing your findings.

% 6. contributions and significance: Describe the key contributions to knowledge (practical, conceptual, theoretical), and how they are of significance (or use to some group of people, organisation, community, etc.).

\end{abstract}





\maketitle

\section{Introduction}

%[Robot in urban space/ challenges]
Robots are evolving beyond their traditional roles in semi-controlled environments, such as industrial and domestic settings, and are increasingly being deployed in more dynamic and unpredictable urban spaces. %, taking on tasks like cleaning~\cite{SALVINI2018UrbanRobotics}, delivery~\cite{FIGLIOZZI2020delivery}, and surveillance~\cite{Halder2023Inspection}.
In these urban spaces primarily designed for human use, robots may encounter situations that extend beyond their pre-programmed abilities, necessitating human intervention for effective operation~\cite{nanavati2021modelingHelpfulness}. This challenge is evident in recent field observations, which reveal instances where urban robots require human assistance for tasks like pressing traffic light buttons~\cite{pelikan2024encountering}, navigating unpredictable obstacles~\cite{Weinberg2023Observe, pelikan2024encountering} and temporarily altered streetscapes~\cite{pelikan2024encountering, Dobrosovestnova2022WithLittleHelp}, or managing conflicts with other road users~\cite{GEHRKE2023Observing}. 

While the primary focus of robotics has long been on developing robots that can autonomously complete tasks without or with little human intervention, the challenges encountered by urban robots highlight that technology alone may not always suffice~\cite{Veloso2018Opportunity}. This has led to increased interest in robots designed to seek human assistance when necessary ~\cite{Cila2022HAC,nanavati2021modelingHelpfulness}.
Furthermore, exploratory projects~\cite{Weiss2010ACE, Smith2017Hitchbot, Tweenbots} and emerging design research perspectives ~\cite{lupetti2019citizenship, Kuijer2018Coperformance,marenko2016animistic} increasingly frame human-robot interaction as a symbiotic relationship, focusing less on a robot's individual capabilities and more on the mutual dependency between humans and robots. 

%[Gap of urban robot help] [seek Bystanders-relationship changes] [Motivation]
Efforts in human-robot collaboration research have been directed towards enabling intuitive and seamless interactions between humans and robots, with humans predominantly cast in the role of collaborative working partners~\cite{ajoudani2018progressHRC}. In urban environments, however, the people urban robots typically encounter and would approach for assistance are often bystanders, or as defined in research,  \emph{`incidentally copresent persons'} (InCoPs)~\cite{Putten2020Forgotten}, people who do not have the deliberate intention of engaging in interactions with the robot. 
This shift in how humans relate to nearby robots calls for the re-examination of existing strategies and the development of new interaction approaches that cater to the spontaneous nature of casual human-robot collaborations in urban settings.

Previous research focusing on the implications of bystander assistance for commercially-deployed urban robots has identified two broad strategies: `helping-as-work', which considers assistance as invisible labour, and `helping-as-care', emphasising the emotional and relational aspects of providing help~\cite{hakli2023helping}.
Extending this discussion, our study introduces a novel dimension: `helping-as-play', which focuses on encouraging bystander assistance through playful interactions. 

Playful strategies and the use of game elements has been found to be effective in various human-robot interaction contexts, from motivating children's learning~\cite{Donnermann2021,chen2023gamified} to enhancing collaboration between factory workers and robots~\cite{Chowdhury2021PlayHRC, Venas2024}. Furthermore, it has been shown to effectively engage the public and create enjoyable experiences among bystanders ~\cite{Lee2019Bubble, Marius2020Woody}. A major challenge for urban robots in soliciting assistance from bystanders lies in persuading them to invest time and manage potential disruptions~\cite{nanavati2021modelingHelpfulness}. Leveraging the inherent playfulness in humans has the potential to entice bystanders into assisting robots by offering a moment of joy within urban environments.

To investigate this opportunity, we developed a series of playful help-seeking design concepts that implement the `helping-as-play' strategy. To test their effectiveness, we compared our playful concepts with concepts that implement the aforementioned strategies: verbal help-seeking, aligning with `helping-as-work' as evidenced by its prevalent use in human-robot teamwork settings~\cite{knepper2015recovering, Srinivasan2016HelpMePlease, Budde2017Needy}, and emotional expression help-seeking, aligning with `helping-as-care' for its effectiveness in eliciting human empathy~\cite{Backhaus2018someone, Zhou2020Sad, urakami2023emotional, Daly2020RobotInNeed}. These strategies were prototyped and evaluated through a virtual reality (VR) experiment involving 24 participants. Our comparative analysis assesses the quality of these help-seeking strategies in terms of their unambiguity, politeness, appropriateness, and effectiveness. Additionally, it examines their impact on people's experiences and moods, as well as their influence on people's attitudes towards the robot.
%Our comparative analysis focused on three key aspects: (1) the assessment of the help-seeking strategies characteristics, (2) the impact on people's experience and mood, and (3) the influence on people's attitudes towards the robot. 
Through our design process and evaluation study, we aim to thoroughly investigate the effectiveness of various help-seeking strategies in eliciting bystander assistance. Additionally, we seek to investigate their broader implications, thus advancing our understanding of how such spontaneous collaborations should be facilitated.

% The investigation is specifically guided by the following research questions: 
% \begin{itemize}

% \item \textit{RQ1: How do different robot help-seeking strategies vary in terms of (1) unambiguity, (2) politeness, (3) appropriateness, and (4) effectiveness?} 

% \item \textit{RQ2: How do different robot help-seeking strategies affect the mood and user experience of bystanders?} 

% \item \textit{RQ3: How do different robot help-seeking strategies influence bystanders' attitudes towards the robot in terms of (1) acceptance, (2) trust, (3) perceived social attributes, and (4) likability? } 

% \end{itemize}

%[Contribution Statement]

This paper makes the following contributions: (1) It introduces playful engagement as a novel strategy for urban robots to seek help from bystanders, along with three exemplary design concepts implementing this strategy; (2) It provides insights into different robot help-seeking strategies and their potential influence on bystanders' experience and attitudes towards robots. This understanding further leads to design considerations of how such robot help-seeking should be facilitated, as well as reflections on the design of game-inspired playful help-seeking strategy. Our study extends the application of playful engagement beyond merely creating enjoyable experiences, demonstrating its effectiveness in supporting the increasingly prevalent casual collaboration between urban robots and bystanders, thus promoting reciprocal co-existence.

\section{Related Work}

\subsection{Service robots in public spaces}
Robots, once predominantly deployed in controlled and semi-contro\-lled configurations such as industrial settings~\cite{Saupp2015Industry} and domestic environments~\cite{Schneiders2021Domestic}, are now expanding their presence into public urban spaces, thereby increasingly becoming integral components of our urban landscapes. These robots provide services in various aspects of society, including sectors such as transportation and logistics, infrastructure maintenance, cleaning, and surveillance~\cite{SALVINI2018UrbanRobot}. 

Despite technological advancements endowing robots with increasing autonomous capabilities, the inherent dynamism and complexity of urban environments pose significant challenges to their operation. This was vividly demonstrated in several viral videos on social media, showing, for example, delivery robots struggling in Estonia's heavy snowfall~\cite{postimees2021robotsnow}. Further, recent field observation studies of urban service robots have also documented instances where their operations were hindered by unexpected obstacles~\cite{Weinberg2023Observe}, the inability to manipulate traffic infrastructure (i.e., traffic light button~\cite{pelikan2024encountering}, human activities occurring on the streets ~\cite{pelikan2024encountering, Babel2022findings}, and bullying behaviours~\cite{ChildrenAbuse2015Brvsvcic, Bully2010Salvini}.

Beyond the evident operational difficulties, these instances intriguingly highlighted the spontaneous and supportive behaviours demonstrated by passersby, showcasing a compelling facet of casually formed human-robot interaction. In the observation study conducted by~\citet{Dobrosovestnova2022WithLittleHelp}, passersby were observed clearing snow in front of the robot or giving it a gentle push to return it to its path. In~\cite{Weinberg2023Observe}, pedestrians voluntarily assisted immobilised robots by removing obstacles. In some cases, people have even interrupted their own activities to assist the robot, as exemplified by a window cleaner pausing their work to allow delivery robots to pass through~\cite{pelikan2024encountering}. These observations underscore the potential of leveraging bystander assistance to enhance the operation of urban robots, resonating with the emerging perspectives in HRI that emphasise relational collaboration over solely technological independence~\cite{Weiss2010ACE, Tweenbots, lupetti2019citizenship}.  

\subsection{Robot help-seeking}
In human-robot collaboration settings, robot help-seeking has been explored as an intentionally designed strategy to recover from inevitable situations that exceed the robot's inherent capabilities. Tested in a human-robot team assembling context, \citet{knepper2015recovering} equipped the robot with verbal communication ability to generate help-seeking requests for tasks like handling unreachable objects. To advance the understanding of verbal help-seeking, research further evaluated factors like ambiguity~\cite{Budde2017Needy} and politeness~\cite{Srinivasan2016HelpMePlease, Budde2017Needy} in framing requests, and their impact on the effectiveness of help-seeking. In addition to explicit and spoken-language help-seeking requests, some studies have successfully tested the use of implicit and non-verbal cues, such as movement~\cite{Kwon2018Incapability}, as well as light and sound ~\cite{Cha2016nonverbal}, to elicit assistance from human collaborators. 
%These methods have also effectively prompted assistive behaviours from human collaborators in joint tasks. 
Furthermore, employing emotional expressions to elicit empathy has been explored as another strategy for encouraging collaborative assistance~\cite{Backhaus2018someone, Zhou2020Sad, urakami2023emotional, Daly2020RobotInNeed} or inducing bystander prosocial interventions during robot abuse~\cite{Abuse2018Tan, prosocial2020Connolly}. For example, research has shown that when robots exhibited sad emotional expressions, people were more inclined to assist them, leading to quicker success in a collaborative game~\cite{Zhou2020Sad}.

Unlike structured human-robot team settings with shared goals between both parties, help-seeking in public spaces poses unique challenges due to misaligned objectives between robot and bystanders. Furthermore, research has shown that diverse factors such as activities bystanders are currently engaged in ~\cite{Huttenrauch2003ToHelpOrNot, Fischer2014SocialFrame, rosenthal2012someone}, the robot's apparent legitimacy and perceived risk~\cite{Piggybacking2017Booth}, and the bystander’s trust in and perceived competence of robots~\cite{Cameron2015Button}, collectively influence their willingness to offer help. However, tailored design strategies or investigations that specifically focus on robot help-seeking from bystanders remain limited.
Some studies have adopted verbal help-seeking~\cite{rosenthal2012someone,Liang2023Direction} which has been predominantly used in traditional human-robot collaboration settings, and a few have explored broader communication modalities, examining the interplay of movement with speech ~\cite{Holm2022Stuck, Fischer2014SocialFrame}.
Thus, the increasing need for assistive behaviors towards robots in public spaces calls for designers to imagine new interaction strategies to engage bystanders in casual collaboration with robots~\cite{Cila2022HAC}.

\subsection{Playful and gameful design in human-robot interaction}
Over the past two decades, technology has become ubiquitous, expanding beyond the context of the workplace. This has marked a significant paradigm shift in the field of interaction design~\cite{harrison2007three}, challenging previous values such as efficiency and placing a stronger focus on meaning-making and enhancing the experiential qualities of interaction. This shift has also resulted in the rise of commercial social robots, as well as numerous examples from research, designed to promote playful engagement while striving for higher-level goals. Examples include playful robots that provide companionship to people in domestic contexts~\cite{Zuckerman2020, Ye2023toaster, Ottoman2015Sirkin}, support children's learning~\cite{Lupetti_2020}, promote creative and critical thinking~\cite{Lee2020Ludic, Lupetti2022}, or trigger social interactions in the context of the city~\cite{Marius2020Woody,Lee2019Bubble, TrashBarrel2023Bu,TrashBin2010Children}. Many of these robotic artefacts are intentionally designed to be open-ended in terms of their form factor and interactions, aiming to foster exploration and self-directed play.

On the other hand, a more structured approach to play encompasses the use of social robots as players or facilitators in games~\cite{mti2040069, Lupetti2016, Zaga2016}, thereby adhering to specific rules and objectives. \citet{mti2040069}, for example, designed and evaluated a mixed-reality playground in which children can play physical games with or against a robot. Furthermore, there have been examples where gameful design elements and principles have been applied to social robots that are deployed in a non-game context, such as education~\cite{Donnermann2021,chen2023gamified, Riedmann2022}, labor~\cite{Chowdhury2021PlayHRC,Venas2024}, and healthcare~\cite{Feingold-Polak2021}. In addition, game elements have been used in robots to promote positive behaviour change in public spaces, such as waste sorting~\cite{Castellano2019Waste}. In the broader interaction design community, the practice of adding gameful design elements and principles to non-game contexts is commonly also referred to as `gamification'~\cite{Tondello2016}. This approach leverages people's intrinsic and/or extrinsic motivation to engage in an activity due to the enjoyment of the task itself or the possibility to attaining a goal~(e.g.,~implemented through a reward system).

%Commonly referred to as ``gamification'', designers make use of ... or applying gameful design principles to the design of robotic application. 
%[TETRABIN]Playfulness in engage people in prosocial interactions~\cite{Marius2018LowRes,tomitsch2014PublicDisplay}
%-- gamification HCI

%-- clarifying terms: playful engagement, games vs play

While previous observation studies have reported on people's playful attitudes towards delivery robots (e.g., ~\cite{Dobrosovestnova2022WithLittleHelp, Weinberg2023Observe, pelikan2024encountering}) and design researchers have documented case studies of urban robotic artefacts designed solely for play~\cite{hoggenmueller2021}, there is a gap on leveraging playful engagement through gameful design to encourage casual bystanders to assist urban robots in situations of failure. Importantly, our research contrasts with most existing implementations of gameful design in HRI, which primarily use robots to engage users in a full game in order to support higher-level goals. Instead, %albeit to achieve higher-level goals, 
we are investigating whether and how gameful design can be used in specific interaction scenarios~(e.g., robot failure), with the robot pursuing its own primary function~(e.g., delivery).
%Importantly, compared to most of the existing applications of gameful design in HRI where the robot's primary functions are , we are investigating whether
%%[Gap: haven't been used in pragmatic context]

\subsection{Summary}
In summary, robots operating in urban environments will inevitably encounter situations that exceed their inherent capabilities, necessitating bystander assistance. 
The involvement of bystanders and the complexity of contextual factors set these casual collaborations apart from traditional human-robot collaborations. This distinction highlights the need for design investigation into strategies that can effectively facilitate such help-seeking in dynamic and unpredictable urban settings. 
Drawing inspiration from playful robots and the use of gameful design elements in robotics to achieve certain goals, we propose `helping-as-play,' a novel strategy that likens robot help-seeking to playful interactions.

%Our work not only extends current views on bystander assistance to urban robots as invisible work or act of care, but also deepens the understanding of various perspectives through the empirical examination of different robot help-seeking approaches. 

%In our study, drawing inspiration from the effectiveness of playfulness in fostering public engagement, we explore `helping-as-play,' a novel concept that likens robot help-seeking to playful interactions.

% Gamification can be defined as the use of game elements in a non-game
% context to engage users and motivate them to adopt specific behaviours~\cite{Gamefulness2011}. It is based on the motivational qualities of good games that make them powerful tools for learning and behavior driving. ~\cite{Michael2017Satisfaction}

\section{Design Concept Development}
In our prior research~\cite{Yu2024HelpSeeking} on encouraging bystander assistance for urban robots, playfulness was identified as a key incentive for fostering bystander help. Building upon these findings, this study advances the exploration of using playful engagement as a strategy for robots to seek help from bystanders. Following an iterative design process, we first conducted a design workshop with members of our research group to engage in collaborative brainstorming and receive feedback on a set of initial concepts. These insights were then incorporated into the final design concepts. 
%This section commences with the description of scenarios in which urban robots require assistance, proceeds with an overview of the design workshop process and the insights gathered from it, and then ends with the introduction of our final design concepts.

\subsection{Help-seeking scenario}
Our design investigation is contextualised in real-world scenarios in which urban robots might encounter operational difficulties. These scenarios were drawn from a comprehensive online ethnography study we previously conducted. In this study, we analysed 177 user-generated videos that captured road users' casual encounters with delivery robots on TikTok~\footnote{https://www.tiktok.com/}. We identified three typical scenarios where an urban robot may face operational difficulties, including (1)~The robot is blocked and needs people to make way for it, (2) The robot is unable to cross the road and needs people to press the traffic light button for it, and (3) The robot is stuck and needs people to push it out. We chose these varied representative scenarios because they encompass different levels of engagement required from bystanders, ranging from simply making way for the robot, to manipulating city infrastructure on the robot's behalf, and to physically pushing the robot when it gets stuck which further requires more effort. At the same time, developing and evaluating design concepts across different scenarios can help validate if the playful help-seeking strategy applies to a broad range of situations that robots may encounter in urban environments.
\begin{figure*}[h]
\begin{center}
\includegraphics[width=1\textwidth]{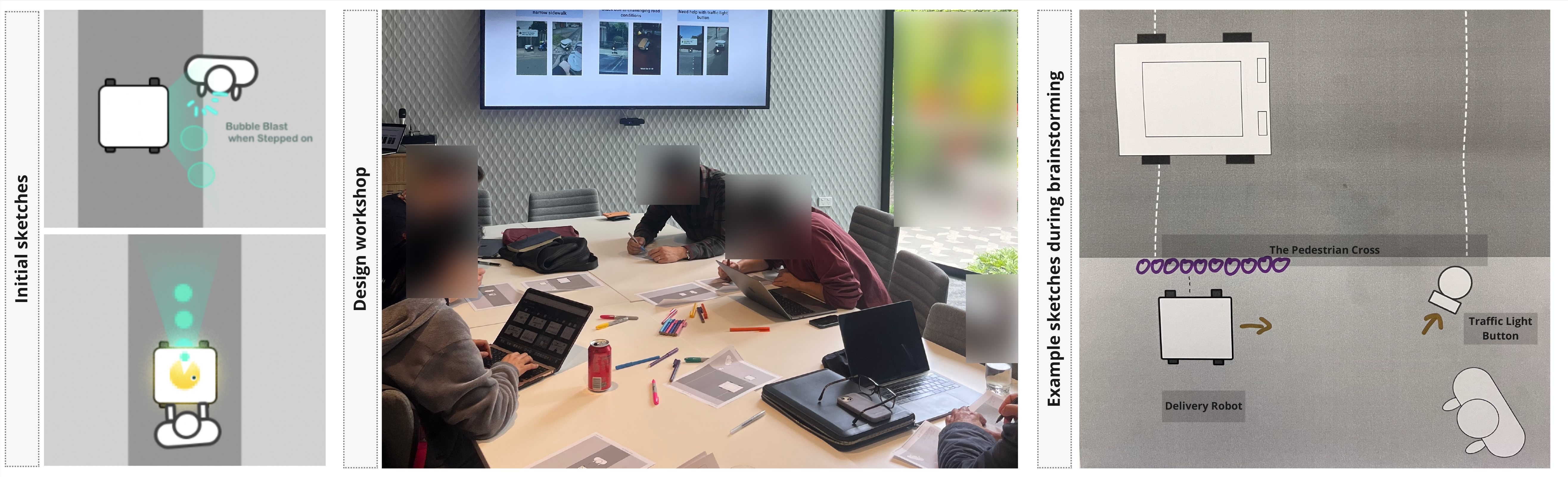}
\end{center}
\caption{The design process. Left: Initial sketches with bubble-blasting (top) and PacMan (bottom) style game elements for promoting helpful behaviours. Middle: Photo from the design workshop brainstorming session. Right: Example sketch generated during the workshop, which was sketched on printed a bird's-eye view illustration of the scenario including the robot, bystanders, and text descriptions.}\label{Process}
\Description{A composite image depicting three stages of the design process. On the left, two initial sketches feature game elements for promoting helpful behaviors: the top sketch shows a bubble-blasting mechanism, and the bottom sketch illustrates a PacMan-style game. The center of the image shows a photo from a design workshop, with several participants actively brainstorming and sketching ideas. On the right, an example sketch generated during the workshop is displayed, presenting a bird's-eye view illustration of an urban scenario with a delivery robot at a pedestrian crosswalk.}
\end{figure*}
\subsection{Design workshop}

The workshop involved five members of our research group (two interaction designers, two engineers, and one urban geographer). It started with a screening of selected TikTok videos, showcasing typical challenges urban robots may face, selected from the dataset of our prior online ethnography study. This provided the workshop participants with a contextual understanding of the scenarios where robots seek assistance. Subsequently, participants were provided with paper representations of the scenarios (see Fig.~\ref{Process} Right) to facilitate brainstorming and idea sketching. They were encouraged to incorporate game-inspired elements with specific rules to promote helping behaviours among bystanders. This focus was chosen over more open-ended playful interactions, as urban robots are functionally oriented and such unstructured interactions could potentially interfere with their operations.
Following the sketching session, all participants presented and commented on each design idea, including the initial design concepts developed by the first author before the workshop.

The ideation session and discussion were consolidated into two key considerations: (1) drawing inspiration from well-known games to ensure intuitiveness in engaging bystanders, and (2) using digital content (i.e. projections) to augment elements in the urban environment. For example, in one of the design ideas generated during the workshop, a traffic light button was reimagined as part of a shooting game (see Fig. \ref{Process} Right). Pedestrians would press it to launch a projectile to break a `brick wall' that blocked the robot's path. (3) Given the functionality-oriented purpose of urban robots operating in urban environments, we decided to make minimal alterations to the robots themselves (e.g., adding lights, transformations). This approach also ensures the applicability of our design concepts across various types of robots.

\begin{figure*}[h]
\begin{center}
\includegraphics[width=1\textwidth]{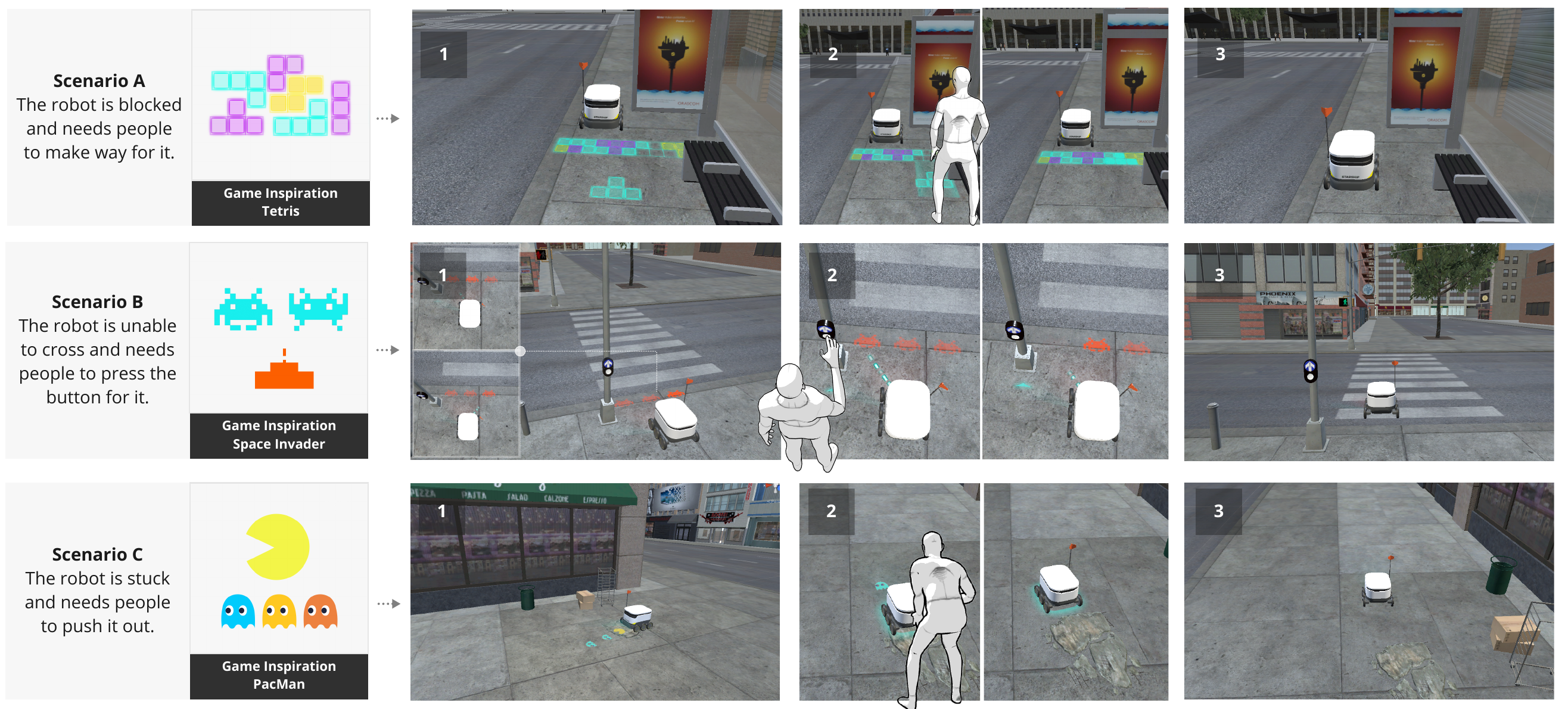}
\end{center}
\caption{The design concepts. Left: Games that inspired the design concepts; Right: Illustration of the process of bystanders engaging in playful helping-seeking. }\label{DesignConcept}
\Description{Figure 2 is a composite image illustrating design concepts for a robot-helping scenario, divided into two main sections. On the left, three panels display the design inspirations for each scenario, each drawing from a classic video game. Design concept for scenario A, inspired by Tetris, shows a robot blocked and needing people to clear the way. Design concept for scenario B, influenced by Space Invader, depicts a robot that requires people to press a crosswalk button. Design concept for scenario C, based on PacMan, illustrates a robot stuck and needing people to push it out. Each panel includes icons inspired by the respective games.

On the right, three sets of three images demonstrate the sequence of bystander interactions in engaging with the three playful help-seeking design concepts in each scenario within a VR simulation. The first set shows a person clearing a path for the robot, the second set displays a person pressing the crosswalk button, and the third set illustrates a person pushing the stuck robot.
}
\end{figure*}

\subsection{Design Concepts}

Building on the ideas and insights generated from the workshop, we further developed final design concepts for robot help-seeking across the three scenarios. For each concept, we drew inspiration from a classic game, utilising its game mechanics to foster an intuitive understanding of the scenarios, motivate bystanders to assist the robot, and elicit interaction.
 
\textit{Scenario A: Robot blocked.}
In this scenario, we incorporated mechanics reminiscent of the retro game `Tetris'. We also drew inspiration from an existing research prototype, TetraBIN, which employed similar game mechanics to encourage people to deposit rubbish into a bin, thereby controlling light blocks on an integrated screen to complete the game~\cite{tomitsch2014PublicDisplay, Marius2018LowRes}. A block is projected in front of the bystander, requiring them to move to control and align the block with a corresponding Tetris-like shape projected in front of the robot (see Fig.\ref{DesignConcept} Top). Once the block and shape match, the block `drops' from the bystander side towards the robot, completing the Tetris line. This action destroys the line in the game, simultaneously creating enough space for the robot to pass through.

\textit{Scenario B: Traffic light button.}
To motivate passersby to press the traffic light button for the robot, we transformed this button into a trigger for a projected shooting game, reminiscent of classic arcade games like `Space Invader' (see Fig.\ref{DesignConcept} Middle). A virtual projectile was projected in front of the robot, continuously rotating to aim at `enemies', accompanied by a projected arrow pointing at the traffic light pole. This arrow serves as an indicator, prompting passersby to press the button. When the robot's perception system detects this button-pressing behaviour, it boosts the projectile, enabling the robot to shoot `enemies'. 

%, which boosts the projectile upon the robot detect people's behaviour of pushing the button through its percepotion system, enabling the robot to shoot the `enemy'. 

\textit{Scenario C: Robot stuck.}
We incorporated mechanics reminiscent of the classic game `Pac-Man', where players collect items to earn points or chase `ghosts' for the bonus. Specifically, we projected several `ghost' images in the intended direction of the stuck robot and a Pac-Man image in front of it (see Fig.\ref{DesignConcept} Bottom). The game mechanics encouraged participants to `chase' the `ghosts' by pushing the robot towards its intended direction, thereby simultaneously aiding the robot out of its stuck position.

\section{Study Design}

%Previous research has categorised bystander assistance for commercially-deployed robots into two broad categories: `helping-as-work', which consider such helping behaviour as invisible labor, and `helping-as-care', emphasising the emotional and relational aspects of providing help~\cite{hakli2023helping}. In our study, we explore `helping-as-play', a novel concept that likens robot help-seeking to playful interactions. 
To gain in-depth insights into our designed playful help-seeking concepts, and to explore their distinct characteristics and those of other robot help-seeking approaches that correspond to the existing perspectives of `helping-as-work' and `helping-as-care', we conducted a within-subject study.
%the playful help-seeking strategy and derived design concepts) with the two existing help-seeking strategies and explore their distinct characteristics, we conducted a within-subject study. 
For `helping-as-work', we opted for a spoken-language robot help-seeking request as previously tested in work environments for task-oriented assistance~\cite{Srinivasan2016HelpMePlease, Budde2017Needy}. For `helping-as-care', we decided on emotional expression as robot help-seeking approach, emphasising the affective and relational aspects of interaction, previously tested in ~\cite{Backhaus2018someone, Zhou2020Sad, urakami2023emotional}.

We decided on an evaluation study in VR that allows participants to experience encounters with and help-seeking requests from a delivery robot in a simulated urban space. VR simulations, now widely used for prototyping and evaluating interactions with robots (e.g., \cite{Xinyan2023Prediction, Milde2023MultiModalVR, VR2020Grzeskowiak,robotics12060168VR}), have been validated for reproducing authentic interaction experiences ~\cite{Villani2018VR, hoggenmueller2021context}. 
Our study focuses on unpredictable scenarios involving urban robots needing assistance, which are difficult to replicate in public spaces. Moreover, conducting the study in VR not only minimises potential risks to participants but also reduces both the implementation time and costs associated with real-world prototypes, thus facilitating a more efficient evaluation method in design research.

%Furthermore, \hlc{our study involves the casually happening unpredictable scenarios where urban robots need help, replicate these scanarios is challenging }{}{}

%VR is particularly suitable for context-specific HRI studies in urban spaces, for \hlc{the difficulties in replicating scenarios in the real world,  high costs of real-world prototypes and potential risks for participants. }{}{}

%which replicating scenarios in the real world would be challenging,~\cite{Villani2018VR}.
%(i.e., the casually happening unpredictable scenarios where urban robots need help). 

Our study is specifically guided by the following research questions:

\begin{itemize}
\item \textit{RQ1: How do different robot help-seeking strategies vary in terms of (1) unambiguity, (2) politeness, (3) appropriateness, and (4) effectiveness?} 

\item \textit{RQ2: How do different robot help-seeking strategies affect the mood and user experience of bystanders?} 

\item \textit{RQ3: How do different robot help-seeking strategies influence bystanders' attitudes towards the robot in terms of (1) acceptance, (2) trust, (3) perceived social attributes, and (4) likability? } 

\end{itemize}

%with (1) speech help-seeking, which we align with the `helping-as-work' as they are commonly used in work environments for task-oriented assistance~\cite{Srinivasan2016HelpMePl ease, Budde2017Needy}, and (2) emotional expression help-seeking, associated with `helping-as-care', emphasising the empathetic and relational aspects of interaction~\cite{Backhaus2018someone, Zhou2020Sad, urakami2023emotional}.

\subsection{Experiment conditions}
In order to compare our playful help-seeking concepts with existing strategies, we conducted a comparative study with three conditions. The conditions mapped respectively to our proposed `helping-as-play,' and the existing strategies `helping-as-work' and `helping-as-care.' In the \emph{Play} condition, we implemented our final design concepts for playful robot help-seeking (introduced in Section 3.3, see Fig.~\ref{DesignConcept}). It is worth noting that our primary objective is to compare these strategies on a conceptual level rather than specific attributes that manifest them, such as audio or visual modalities.

For the \emph{Work} condition, we aligned the verbal help-seeking approach with `helping-as-work' strategy, as it is commonly employed in the context of human-robot teamwork~\cite{knepper2015recovering, Srinivasan2016HelpMePlease, Budde2017Needy}. To compose the verbal help-seeking content, we followed the structure of \emph{Justification} (i.e., interpreting the situation regarding the help needed) + \emph{Introduction} (i.e., communicating a definitive instruction on how an individual should provide help) as proposed in~\cite{Budde2017Needy} and has also been employed in the setting where a delivery robot asks bystanders for help~\cite{Boos2022Delivery} (see Table~\ref{Speech} for detail request content). The robot's help request was recorded using the `Alex' voice of the macOS text-to-speech engine, as per the method in ~\cite{Srinivasan2016HelpMePlease}.

\begin{table}[]
    \centering
    \small
    
        \caption{Speech help-seeking in each scenario}
        \Description{The table titled 'Speech help-seeking in each scenario' lists verbal help-seeking strategies used by a robot in three different scenarios. Each scenario is outlined with two columns: Introduction and Justification. In Scenario A, labeled 'Robot Blocked', the robot's introduction is 'Please let me pass' with the justification 'I am going to be late'. Scenario B, titled 'Traffic Light', has the introduction 'Please help me press the traffic light button' with the justification 'I can't reach it'. Scenario C, 'Robot Stuck', includes the introduction 'Please push me out of this spot' and the justification 'I am stuck'. 
}
        \label{Speech}
        \begin{tabular}{p{2.1cm}p{2.7cm}p{2.5cm}}
            \toprule
            \textbf{Scenario} & \textbf{Introduction} & \textbf{Justification} \\
            \midrule
            A: Robot Blocked & Please let me pass  & I am going to be late \\
            B: Traffic Light & Please help me press the traffic light button& I can't reach it \\
            C: Robot Stuck & Please push me out of this spot & I am stuck  \\

            \bottomrule   
         
        \end{tabular}

\end{table}

\begin{figure}[h]
\begin{center}
\includegraphics[width=0.5\textwidth]{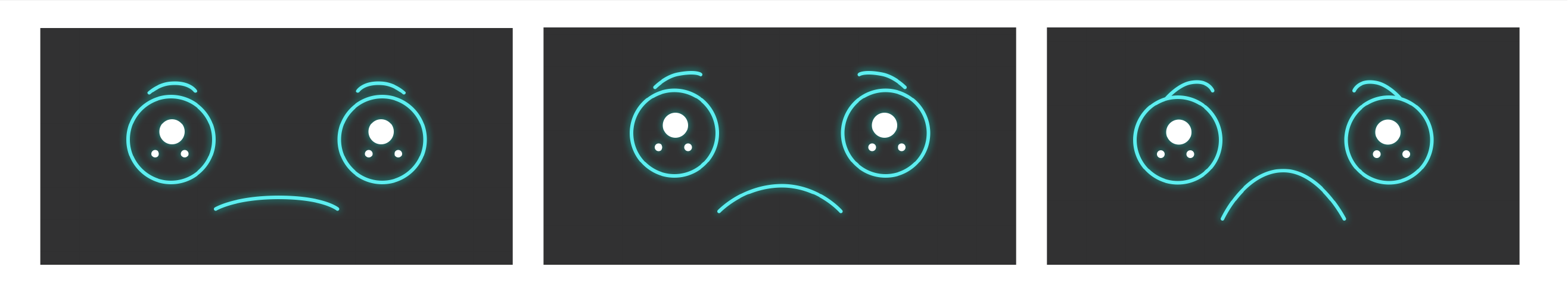}
\end{center}
\caption{Key frames of robot facial expression}\label{face}
\Description{Key frames of robot facial expression. The image displays three consecutive frames of a robot's face designed to show different expressions. From left to right: the first frame shows the robot with a neutral expression, characterised by a flat mouth and circular eyes. The second frame shows a slightly saddened expression, with the mouth turned downward. The third frame retains the sad expression but with a slightly more pronounced downturn of the mouth. All expressions are depicted using simple line drawings on a black background, emphasizing the changes in the mouth to convey different emotions.
}
\end{figure}

In the \emph{Care} condition, we aligned emotional expression help-seeking with `helping-as-care' strategy due to its effectiveness in eliciting human empathy, thus encouraging assistive behaviours~\cite{Backhaus2018someone, Zhou2020Sad, urakami2023emotional, Daly2020RobotInNeed}. We implemented sad facial expressions, as existing research suggests that robots displaying sad emotional expressions can elicit an increased willingness to help from people~\cite{van2020designing,hakli2023helping, Herdel2021Drone}. We used the same sad facial expression as in ~\cite{Herdel2021Drone}, because of its proven effectiveness in eliciting prosocial responses from bystanders in similar casual encounter settings. Following their approach, we created a set of faces as key-frames (See example key-frames in Fig.~\ref{face}) and blended them into an animation sequence.

%with (1) speech help-seeking and (2) emotional expressions help-seeking. We align `helping-as-work' with speech help-seeking, as these are typically used in work environments and focus on task-oriented assistance~\cite{Srinivasan2016HelpMePlease, Budde2017Needy}. For `helping-as-care', we consider emotional expression help-seeking strategies~\cite{Backhaus2018someone, Zhou2020Sad, urakami2023emotional}, which underscore the empathetic and relational dimensions of interaction. 

\subsection{Study apparatus and implementation}
We simulated the three scenarios in Virtual Reality using Unity3D\footnote{https://unity.com/}, implementing the three different help-seeking strategies for a delivery robot for each scenario. We used off-the-shelf 3D models from the Unity Asset Store to construct scenarios, such as sidewalks and pedestrian crossings with traffic lights. For the mobile robot, we used a 3D model of the delivery robot Starship, obtained from the modeling platform Sketch Fab\footnote{https://sketchfab.com/}. We employed the HTC Vive headset\footnote{https://www.vive.com/} for participants to engage in the various robot help-seeking scenarios. This setup enabled them to move freely within a 3m x 3m tracked area and interact with the virtual objects (e.g., traffic light button, robot) using simulated hands by holding the controllers.

\subsection{Participants}

We recruited a total of 24 participants in the age range of 18 to 74 years. The majority (n=17) were between 25 and 34 years old, with three aged 35-44 and two aged 18-24. Our participant cohort also included two representatives from older demographics, including one participant each from the age groups of 55-64 and 65-74, respectively. Thirteen of our participants self-identified as female, ten as male, and one preferred not to disclose their gender. Participants were recruited from our university’s mailing lists, flyers and social networks. All participants voluntarily took part in the experiment and initial contact had to be made by them, following the study protocol approved by our university’s human research ethics committee. 

\subsection{Procedure}

After participants arrived at the study site, they were first given a brief introduction about the study background and procedure. They were then asked to sign a consent form, followed by a demographic questionnaire that obtained their basic information, including age group, gender, occupation, nationality, and previous experience with AR/VR and robots. Following this, we briefly introduced the VR headset and its basic operations, and notified participants that the HMD VR was used to simulate the experience of interacting with urban robots.

Before the experiment commenced, each participant went through a familiarisation session to practice walking and interacting with the controllers in the VR environment (i.e., push a rack on wheels). We ensured that participants did not experience motion sickness and were willing to proceed with the study. By the end of the familiarisation session, participants were asked to fill out a single-item questionnaire that measured their mood.

During the experiment, each participant experienced the three experimental conditions in a different order (counterbalanced using a balanced Latin Square design which one condition precedes another exactly twice~\cite{bradley1958complete}) to minimise carryover effects. %, namely \emph{Work} for speech help-seeking, \emph{Care} for emotional expression help-seeking, and \emph{Play} for our design concepts of playful help-seeking. 
%All participants experienced them in a different order (counterbalanced using a Latin Square design) to minimise carryover effects.
Each experimental condition encompassed all three introduced scenarios: \emph{Robot blocked}, where the robot is blocked by participants and needs them to make way for it; \emph{Traffic light}, where the robot requires participants to press the traffic light button for it; and \emph{Robot stuck}, where the robot is immobilised and needs participants to help it out of its stuck position. The order in which each participant experienced the scenarios was also randomised.

Before each scenario, participants were provided with a context to aid their immersion, such as waiting for a bus at a bus station. Participants were not informed about the robot's intention to seek help before the study; they were simply instructed to respond spontaneously to any cues from the robot. This aimed to reduce any sense of obligation to assist the robot resulting from the study's purpose. Each scenario concluded with participants either providing help to the robot or indicating their refusal to engage in further interaction. Subsequent to each scenario, participants filled out two single-item questionnaires that measured their mood and willingness to assist the robot through an in-headset interface. Upon completing all three scenarios within each condition, participants removed the headset and were requested to fill out several standardised questionnaires that assessed help-seeking quality, user experience, and their perceptions of the robot. After participants completed all three conditions, we conducted a post-study semi-structured interview to gain in-depth insights into their experiences. The whole study took approximately 60 minutes.

\begin{figure*}[h]
\begin{center}
\includegraphics[width=1\textwidth]{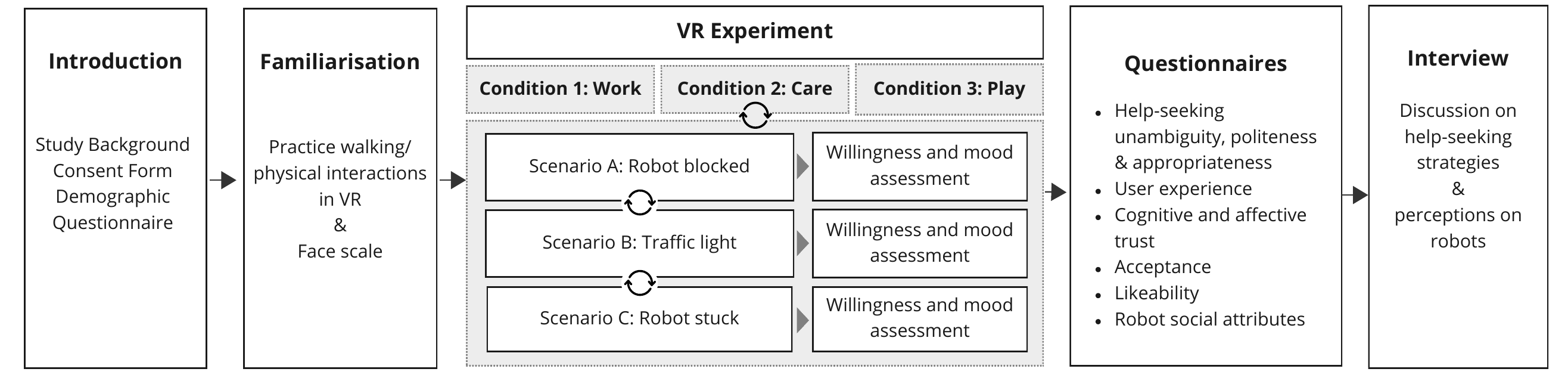}
\end{center}
\caption{Study procedure}\label{Procedure}
\Description{Figure 4: Study procedure. The flowchart outlines five main stages of the study process. Starting with "Introduction," where study background information is provided, consent forms are signed, and demographic questionnaires are completed. Next is "Familiarisation," where participants practice walking and physical interactions in a virtual reality (VR) setting, including using a face scale. The third stage is the "VR Experiment," divided into three conditions: Work, Care, and Play. Under each condition, three scenarios are tested: Robot blocked, Traffic light, and Robot stuck, where participants' willingness and mood are assessed. The fourth stage is "Questionnaires," covering help-seeking appropriateness, user experience, trust, acceptance, likeability, and social attributes of robots. The final stage is "Interview," discussing help-seeking strategies and perceptions of robots.
}
\end{figure*}

\subsection{Data Collection}

We collected both quantitative and qualitative data through questionnaires, observations, and interviews, following a mixed-methods approach ~\cite{Creswell2014MixMethod}. The data collection was structured to address each of the three research questions correspondingly.

\textit{Help-seeking quality:} To assess the quality of help-seeking strategies, we used items from previous HRI studies investigating robot help-seeking to measure the perceived ambiguity, politeness ~\cite{Budde2017Needy} and appropriateness~\cite{Srinivasan2016HelpMePlease}. In addition, we assessed participants' willingness to help the robot after each scenario using the same item as in~\cite{Budde2017Needy}. Each dimension was rated on a 7-point Likert scale ranging from 1 to 7.

\textit{Experience and mood:} To measure participants' experience of their interaction with the urban robot within the simulated VR environment, we used the short version of the widely adopted User Experience Questionnaire (UEQ-S)~\cite{Schrepp2017UEQS} on a 7-stage scale from -3 to +3. 
Helper's high refers to the phenomenon that helping someone or something else can lead to psychological benefits such as mood improvement~\cite{KWAK2018helper}. To investigate its relevance in the context of helping public space robots, we also assessed participants' moods before the study and after each scenario. We used the face scale developed by~\cite{lorish1986face}, which was previously used in another HRI study~\cite{Chirapornchai2021HelperSHigh} to evaluate mood changes after helping a robot. The scale features faces numbered 1 to 20 in descending mood order, with 1 indicating the most positive and 20 the most negative mood. It is worth noting that mood assessment results from instances where participants did not offer help were excluded from our analysis, as our focus was on assessing mood changes after helping behaviour. 

\textit{Attitudes towards the robot:}
We used the Robotic Social Attributes Scale (RoSAS)~\cite{Carpinella2017RoSAS} to measure participants’ perceptions of the robot, which comprises three factors: warmth, competence, and discomfort. In addition, we used the likeability subscale from the Godspeed Questionnaire~\cite{bartneck2009GodSpeed} to measure the perceived likeability of robots. For evaluating trust, we utilised the trust scale developed by~\citet{McAllister1995AffectAC}, which has been employed in assessing human trust towards robots~\cite{zieger2023happiness}. This scale consists of two subscales: cognitive trust and affective trust. We selected this scale over other trust scales~\cite{jian2000Trust, malle2023trust} as its two subscales help differentiate the effects of help-seeking strategies for two types of trust: trust based on rational assessment of the robot's capabilities and trust based on emotional connections to the robot. This differentiation allows for a more nuanced understanding of how various help-seeking strategies influence trust-building between bystanders and robots. %This allowed us to differentiate the impact of various help-seeking strategies on trust, such as whether it was based on a rational assessment of the robot's abilities or the feeling of emotional connection towards the robot.
Lastly, to assess the impact of robot help-seeking strategies on participants' acceptance towards urban robots, the System Acceptance Scale developed by ~\citet{VANDERLAAN19971Acceptance} was used. Each measurement was rated on a 7-point Likert scale ranging from 1 to 7.

\textit{Semi-structured interviews}:
% The Multimodal Presence Scale (MPS) ~\cite{MAKRANSKY2017MultimodalPresenceScale}, which measures physical, social, and self-presence, was employed to evaluate the effectiveness of VR in simulating real-world encounters and interactions with urban robots.
%We took observation notes of participants' reactions towards the robot by observing their physical behaviour and monitoring interactions in the virtual environment from the PC running the simulation. After each experimental conditions, we asked participants to debrief what happened in the scenarios and why they decided to help or not to help the robot. 
Following each experimental condition, participants were asked to provide brief feedback on their experiences in the scenarios, including their reasons for deciding to assist or not assist the robot. After finishing all experimental conditions, we conducted a semi-structured interview with participants regarding their preferences for and opinions on the different strategies and design concepts. The aim was to gain an in-depth understanding of how the various strategies influenced their willingness to offer help and shaped their overall perceptions of the robots.

\subsection{Data Analysis}

{\emph{Questionnaires:}} We first assessed the internal reliability of all multi-item scales by calculating Cronbach’s alpha.  The internal reliability for the help-seeking assessments of \emph{unambiguity} ($\alpha$=0.864) and \emph{appropriateness} ($\alpha$=0.804) was found to be good. The UEQ-S scales showed good reliability for \emph{pragmatic} quality ($\alpha$=0.815) and excellent reliability for \emph{hedonic} quality ($\alpha$=0.928). The system acceptance scale demonstrated excellent reliability with $\alpha$=0.918. For the RoSAS, the \emph{competence} and \emph{warmth} subscales showed good ($\alpha$=0.899) and excellent ($\alpha$=0.900) reliability, respectively, while the \emph{discomfort} subscale had acceptable reliability of $\alpha$=0.771. Lastly, the likability assessment of the robot received excellent reliability with an $\alpha$ value of 0.948. 

We then proceeded to conduct descriptive and inferential analyses of the questionnaire data. Given the non-parametric nature of the data, we used the Friedman test to identify any statistically significant differences. In cases where significant differences were found, we performed pairwise comparisons with Bonferroni corrections. A p-value of less than 0.05 was considered indicative of a significant effect.

{\emph{Interviews:}} All interviews were transcribed by the first author, employing a mixed thematic analysis approach ~\cite{braun2006thematic}.  First, all interviews were coded inductively and initial codes were sorted into sub-themes. The sub-themes were then deductively grouped into final themes, structured around the three main aspects of our research questions, aligning with the Results section's organisation.

\section{Results}

\subsection{Received help}

In our study, each participant experienced nine instances of help-seeking (3 conditions x 3 scenarios), totaling 216 help-seeking instances. The robot received help in the majority of these instances. Exceptions where participants refused to help were few: 
In \emph{`Work'} condition, p18 and p8 did not help in two instances where the robot got stuck, due to the required physical effort to push the robot. In \emph{`Care'} condition, p10, p4, and p17 did not assist in three instances, with two involving a stuck robot and one in the traffic light button scenario. In \emph{`Play'}, p1 did not provide assistance in two cases, one with the robot stuck and another involving pressing the traffic light button. The lack of assistance in both the \emph{`Care'} and \emph{`Play'} conditions was attributed to participants not fully understanding the robot's help request.

\subsection{Help-seeking assessments: unambiguity, politeness, appropriateness, and effectiveness}
\subsubsection{Quantitative results}

\begin{figure*}[h]
\begin{center}
\includegraphics[width=1\textwidth]{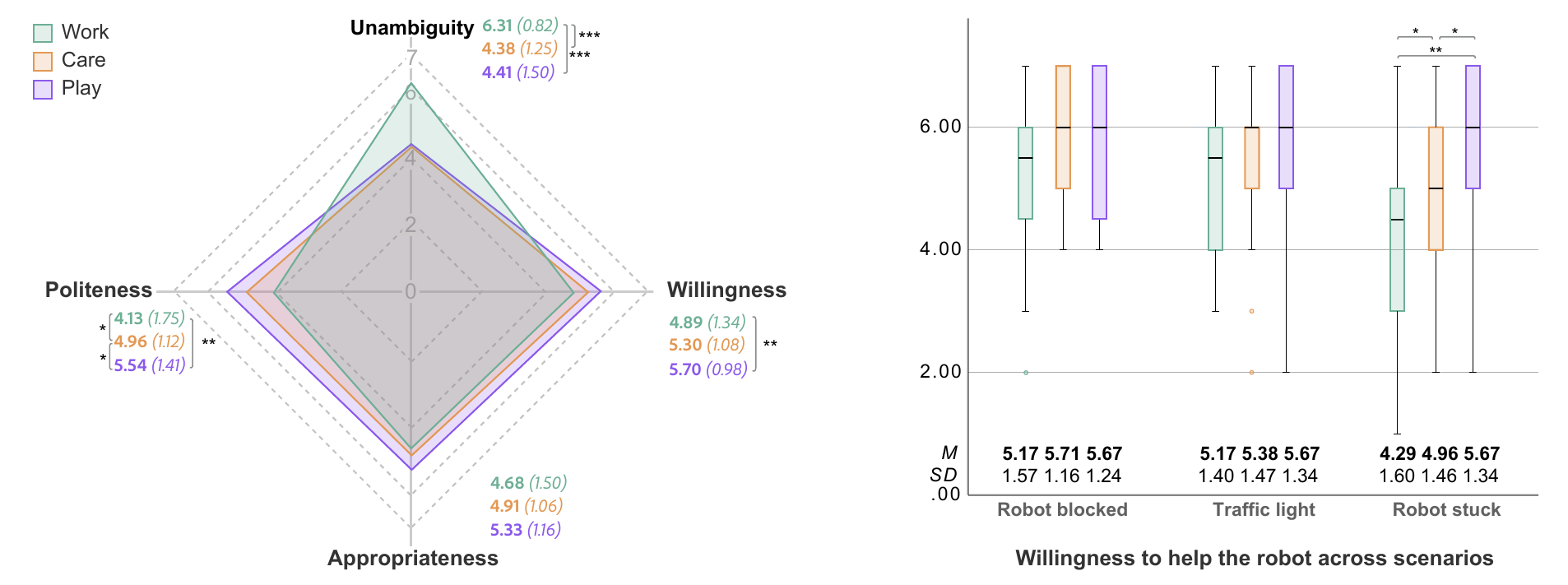}
\end{center}
\caption{Qualitative assessment of help-seeking strategy (Left), Box plot of participants' Willingness to Help the robot across each scenario (Right). M: Mean, SD: Standard Deviations, *: p < .05, **: p < .01, ***: p < .001}\label{Quality}
\Description{Figure 5 is composed of two graphical representations assessing participant responses in a study on help-seeking strategies. On the left, a radar chart displays qualitative assessments for three strategies: Work, Care, and Play, each evaluated on Unambiguity, Politeness, and Appropriateness. Scores are given as follows: Work scores 4.32 on Unambiguity, 4.13 on Politeness, and 4.59 on Appropriateness; Care scores 4.41, 4.96, and 5.31; and Play scores 4.50, 4.54, and 5.33 respectively. Care and Play strategies rate higher on Unambiguity and Appropriateness, while Work is highest on Politeness.

On the right, a box plot illustrates participants' willingness to help the robot across three scenarios: Robot Blocked, Traffic Light, and Robot Stuck. Willingness to help is numerically represented with means and standard deviations: Robot Blocked at 5.17 (SD 1.57), Traffic Light at 5.67 (SD 1.24), and Robot Stuck at 4.96 (SD 1.60). Statistically significant differences in willingness are noted, with the highest willingness observed in the Traffic Light scenario and significantly lower willingness in the Robot Stuck scenario. 

}
\end{figure*}

In the descriptive analysis (see Fig.~\ref{Quality}), help-seeking unambiguity was highest in the \emph{Work}, whereas assessments of politeness and appropriateness were notably the lowest in the same condition. \emph{Care} and \emph{Play} showed similar levels of unambiguity, with \emph{Play} receiving the highest scores in both politeness and appropriateness. Friedman's ANOVA revealed significant differences in both unambiguity ($\chi^2(3)$=27.179, p<0.001) and politeness ($\chi^2(3)$=6.659, p=0.036), but not in appropriateness ($\chi^2(3)$=5.37, p=0.068). 
Post-hoc tests revealed that \emph{Work} received significantly higher unambiguity ratings compared to
both \emph{Care} ($p<0.001$) and \emph{Play} ($p<0.001$), with no significant differences between \emph{Care} and \emph{Play}. Conversely, participants' politeness ratings in \emph{Work} were significantly lower than in \emph{Care} ($p=0.04$) and \emph{Play} ($p=0.002$), with \emph{Play} further exhibiting a significantly higher politeness score than \emph{Care}~($p=0.03$).

Participants' willingness to help the robot was highest in \emph{Play}, and was lowest in \emph{Work}. Friedman's ANOVA revealed significant differences ($\chi^2(3)$=8.47, p=0.015). Subsequent post-hoc tests revealed a significant increase in willingness to help only when comparing  \emph{Play} to \emph{Work} ($p=0.004$). We further analysed participants' willingness to help the robot across different scenarios. The results showed that the willingness ranking remained consistent with the overall assessment for both the robot stuck and traffic light scenarios. However, in the robot block scenario, willingness to help was slightly higher in \emph{Care} than in \emph{Play}. Friedman's ANOVA revealed significant differences only in the robot stuck scenario, which requires the most physical effort for participants to push the robot out of the stuck,  ($\chi^2(3)$=7.136, p=0.028). Post-hoc tests revealed that participants exhibited a significantly higher willingness to help the robot in \emph{Play} compared to both \emph{Care} ($p=0.03$) and \emph{Work} ($p=0.002$), with \emph{Care} also demonstrating significantly higher willingness than \emph{Work} ($p=0.03$). 

\subsubsection{Qualitative feedback on help-seeking quality}
\ 
\newline
\textit{Unambiguity:}
Most participants (n=21) perceived verbal help-seeking as clear and straightforward, while some considered the helping-seeking strategies used in \emph{Care} (n=10) and in \emph{Play} (n=12) as lacking in unambiguity, thus necessitating closer observation of the situation to understand the robot's requests (n=7). The unambiguity and efficiency of speech communication were indicated by five participants as the primary reason for rating verbal help-seeking as their most preferred strategy. Seven participants also found the playful help-seeking approach easy to understand. Some attributed this to recognising the game from which the help-seeking originated (n=3), while others mentioned cues from the visualisation that helped infer the robot's intention~(n=7), such as the `Pac-Man' projection indicating the robot's intended direction. P7, who belonged to the older age group of over 65 and was not acquainted with some of the games, managed to engage successfully, albeit after a brief period of contemplation, as expressed in their remark, \emph{`it did take me a minute to think'}. This learning curve in grasping the help-seeking request was also mentioned by the other nine participants, with seven of them stating that their comprehension was developing during the interactive process, such as seeing \emph{`projections, kind of responding to my movements.'} (p12) 

Regarding emotional help-seeking, nine participants easily inferred the robot's need for help from its sad face, yet were unsure of the specific assistance required, as p3 highlighted: \emph{`[...] you know that it needs help, but it's not clear that it wants to cross the road and it wants you to press the button.'} Furthermore, participants indicated several instances of misunderstanding in both \emph{Care} (n=4) and \emph{Play} (n=7) conditions. For instance, some misconstrued the projection in \emph{Play} as merely \emph{`advertisements'} (p5, p13), while others misinterpreted the robot's sad facial expression as being \emph{`out of battery'} (p4) or \emph{`malfunctioning'} (p5).  

\textit{Politeness:}
Rudeness and impoliteness were prominent in participants' comments about verbal help-seeking (n=16), with p3 and p8 notably exclaiming \emph{`How rude!'} immediately after hearing the robot's request during the study. This aversion impression can be attributed to participants' perception of verbal help-seeking as resembling \emph{`commands'} (n=8), being \emph{`demanding'}~(n=3), or giving them the feeling of being \emph{`ordered around'} (n=6). The impression of an authoritative directive further led seven participants to express that they were not helping the robot out of their own will but felt obligated and even having a feeling of being \emph{`forced'} (p2, p9, p11). P16's statement exemplified this sentiment: {`I still feel like I have to help it even though I didn't like helping it as much. [...] being given the instruction, it almost feels like you have to help it, kind of thing.'}  P11 even expressed concern that the robot might harm them if they did not comply with its requests, stating, {`Maybe if I did not follow the sound commands, it would hurt me.'} In contrast, participants commented that \emph{Care} (n=4) and \emph{Play} (n=3) help-seeking strategies were less assertive and demanding. This perception could stem from the feeling that they have an option not to engage if they lack interest in these two conditions (n=9). For instance, as p1 indicated, \emph{`If you don't have the desire to help, you can just walk away. [...] It's not demanding you to do anything.'}

\textit{Appropriateness:}
Furthermore, five participants found the robot's verbal help-seeking abrupt or unexpected, as p5 indicated:\emph {`I was a bit surprised when it first spoke to me [...] you wouldn't expect a robot to randomly ask a stranger for help.} While the verbal help-seeking approach is direct, it was also perceived as interruptive by four participants. In contrast, the more implicit, playful strategies (n=6) and emotional expressions (n=2), though lacking unambiguity, were deemed less invasive and less likely to cause interruptions. 

Five participants expressed feelings of aversion towards the robot verbally asking for help. P14 highlighted this sentiment: \emph{`I feel like it is blaming me for being in his way'.} Four participants conveyed their discomfort with the robot displaying negative emotions in the \emph{`Care'} condition. This sentiment was echoed by p17, who stated, \emph{`I don't think anyone should be responsible for their [the robots'] personal negative energy.'} Moreover, P17 elaborated that their decision to help the robot was because they \emph{`just wanted to finish this [the interaction]. I don't want to see that sad face.'}

\textit{Willingness to help the robot}
For the factors motivating participants, eleven participants expressed a neutral sentiment in \emph{Work} condition. They did not cite specific reasons for their willingness to help, suggesting it was a natural response for them, especially if it didn't require \emph{`too much effort'} (p1) or if they were not \emph{`in a rush.'}(p7, p17) In the \emph{Care} condition, the predominant factor motivating participants to offer help was the empathy evoked by the robot's sad face, as indicated by nine participants. Regarding the playful strategies, four participants indicated that the projections sparked their curiosity to engage further. This intrigue was indicated by p18, who expressed the intention to \emph{`take on a challenge'} and \emph{`to see if I could work it out'}. Furthermore, some participants also raised concerns about the effectiveness of help-seeking strategies for long-term deployment. Four participants indicated that they might not help the robot again in the \emph{Work} scenario if they encountered it frequently in their daily lives. As p3 noted, they could \emph{`accept it (verbal help-seeking) [...] if it's a one-time or two-time thing,'} but would be reluctant to offer help if they {`encounter this every day on the road'}. Conversely, four participants indicated their willingness to assist the robot again if it used playful help-seeking strategies, as p9 noted it made them \emph{`delighted and more willing to help again'.}

% Conversely, four participants suggested they would be willing to assist the robot using playful help-seeking again

It is noteworthy that in scenarios where the robot was stuck and needed people to push it, six participants reported a decreased willingness to help due to the increased physical effort involved. However, p14 and p21 highlighted that the playful strategies helped to mitigate the perceived exertion involved in assisting the robot, as p14 suggested: \emph{` distracting me from the process of helping it'}. This observation aligns with quantitative results, showing that participants' enhanced willingness to help robots in the \emph{Play} condition was more pronounced in the robot stuck scenario than in scenarios requiring less physical involvement (See Fig.~\ref{Quality}, Right).

\subsection{Participant experience and mood change}
\begin{figure*}[h]
\begin{center}
\includegraphics[width=1\textwidth]{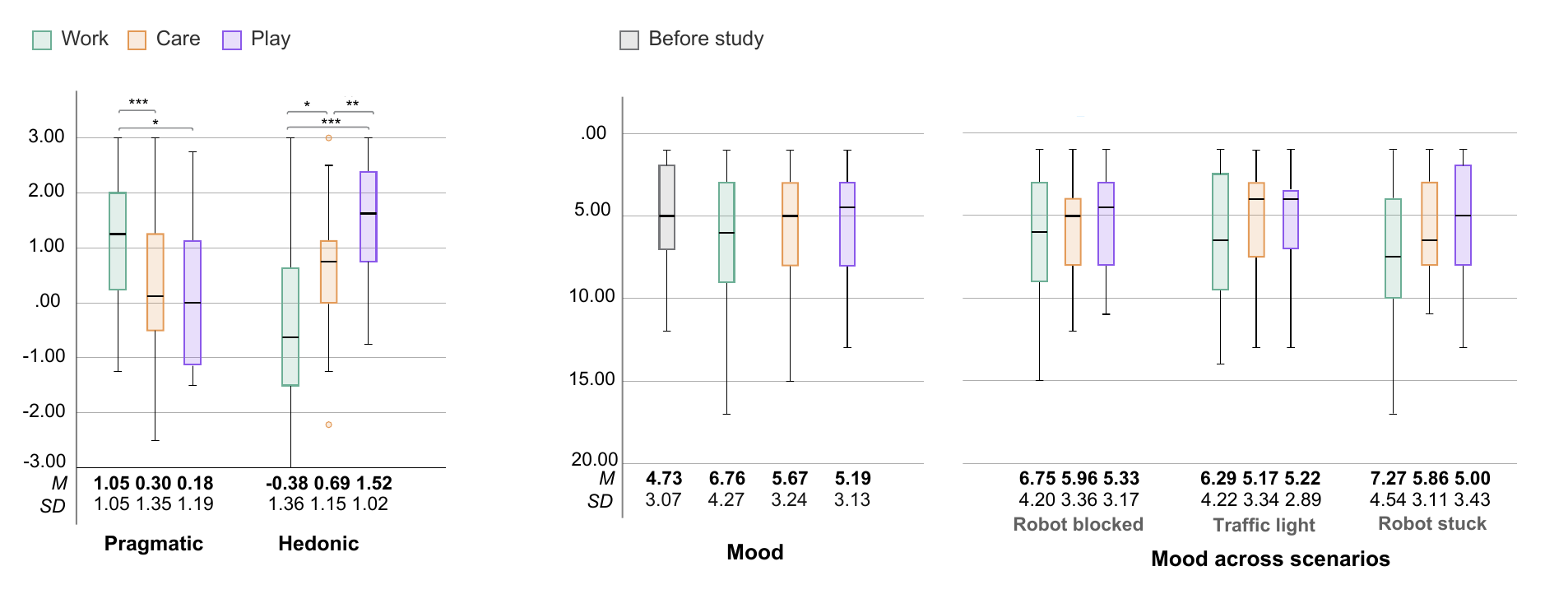}
\end{center}
\caption{Box plot of UEQ results (Left), and mood assessment before and after each scenario (Right) M: Mean, SD: Standard Deviations, *: p < .05, **: p < .01, ***: p < .001}\label{experience}
\Description{Figure 6 contains two sets of box plots. On the left, the box plots depict the results of the User Experience Questionnaire (UEQ) across two dimensions: Pragmatic and Hedonic, analyzed for three strategies: Work (gray), Care (green), and Play (purple). The Pragmatic dimension scores are Work at 1.05 (SD 0.31), Care at -0.38 (SD 0.69), and Play at -1.36 (SD 1.52). The Hedonic dimension scores are Work at 1.35 (SD 1.18), Care at 1.36 (SD 1.50), and Play at 1.05 (SD 1.02). Statistical significance is noted between Work and Play in both dimensions, with Work scoring significantly higher.

On the right, two sets of box plots represent participants' mood assessments across three scenarios—Robot Blocked, Traffic Light, and Robot Stuck—both before and after each scenario (colored in gray for before and orange, green, and purple respectively for after). Mood is measured on a scale from 0 to 20, with means shown for each: Before study, mood averages around 4.73 to 6.76, while after, mood varies more significantly, especially after the Traffic Light scenario where mood improves to 6.33 from 4.20 before. 
    }
\end{figure*}
\subsubsection{Quantitative Results}

Descriptive data analysis (see Fig.~\ref{experience}) of UEQ ratings revealed \emph{Work} as highest in pragmatic but lowest in hedonic quality, with \emph{Play} exhibiting the inverse pattern. Friedman's ANOVA showed significant differences in both pragmatic ($\chi^2(3)$=8.25, p=0.016) and hedonic quality ($\chi^2(3)$=16.795, p<0.001). Post-hoc tests revealed that \emph{Work} received significantly higher ratings for pragmatic quality than both \emph{Care} ($p<0.001$) and \emph{Play} ($p=0.03$), with no significant differences between \emph{Care} and \emph{Play}. Hedonic ratings were significantly higher in \emph{Play} compared to both \emph{Work} ($p<0.001$) and \emph{Care} ($p=0.0013$), with \emph{Care} also scoring significantly higher than \emph{Work} ($p=0.03$). 

Participants reported better mood levels before the experiment than after assisting the robot in all three conditions, with the worst mood assessment after assisting the robot in the \emph{Work} condition. Participants' moods after helping the robot in different scenarios showed similar patterns to the overall mood assessment, with \emph{work} consistently rated the worst across all three scenarios. \emph{Play} was rated higher than \emph{Care} in robot stuck and robot block scenarios, but slightly lower in the robot traffic light scenario. 
Friedman's ANOVA showed no significant differences in the overall mood assessment, nor in comparisons within each scenario. 
%Friedman's ANOVA revealed significant differences only in the scenario where the robot stuck and needs people to push it out ($\chi^2(3)$=6.095, p=0.047). Post-hoc tests indicated that participants' moods after helping the robot in both \emph{Play} ($p=0.003$) and \emph{Care} \textbf{($p=0.01$)} were significantly better than in \emph{Work}, with no significant differences between \emph{Care} and \emph{Play}. 
%Mood assessment results from instances where participants did not offer help were excluded from our analysis, as our focus was on assessing mood changes after helping behaviour. 

\subsubsection{Qualitative feedback: }
 In the \emph{Work} condition, participants generally appreciated the efficiency of the speech strategies used for requesting help (n=16), even though it's \emph{`less interesting'} compared to the other two approaches (n=5). Four participants reported a neutral feeling after helping the robot, with P1 indicating, \emph{`It feels like I'm just fulfilling a task, like doing a job'}. Two participants found speech help-seeking {`less rewarding'} compared to the other two methods, and five participants reported feeling unpleasant after assisting the robot. For example, p21 described a decline in mood after helping the verbally help-seeking robot, stating their \emph{`mood went down'} after having to repeatedly help the robot three times. Five participants attributed their neutral or unhappy feelings to the robot's lack of response after assistance. P14 exemplified this sentiment: \emph{`I was expecting it to say thank you or something like that, but it doesn't provide any feedback, which makes me a bit unhappy'}. Furthermore, six participants expressed concerns about the accessibility of the auditory help-seeking approach, worrying that it might not be heard in noisy urban environments or by people wearing headsets.

 For the \emph{Care} condition, eight participants reported positive feelings after assisting the robot. Four of these participants noted that their positive response was similar to the feeling of helping other people or small animals in need. P20 encapsulated this sentiment: \emph{`I would say that I felt the best when I helped the one with the sad face because it felt like I was, [...] helping out a person, almost.'} However, due to the robot's human or animal-like expressions, more participants -- compared to the \emph{Work} condition -- expected a response (n=9) or wished for more interactions with the robot (n=3). For example, some participants expressed a wish to see positive reactions, such as a smiling face (n=5) or behaviours indicating excitement, like a \emph{`dance to show some excitement'} as described by p17. Five participants reported negative feelings due to the lack of response from the robot after assisting it. Furthermore, five participants noted difficulties in seeing the robot's face when standing at the side of the robot or when pushing the robot from behind, factors which could negatively impact their experience.

Participants generally feel that their experience of helping the robot in \emph{Play} was \emph{`fun'} (n=8), \emph{`enjoyable} (n=8), and \emph{`interesting'} (n=6). Additionally, two participants expressed excitement due to the novelty of the approach. However, p14 also noted that the sense of \emph{`freshness will disappear'} with long-term deployment. Furthermore, in the \emph{Play} condition, half of the participants expressed a \emph{`sense of accomplishment'} or achievement following their assistance to the robot. This feeling was exemplified by p15, who reflected, \emph{`I felt more fulfilled, like, oh, I did something.'} Six participants noted a disconnect between game-playing and assisting a robot, with p16 describing it as \emph{`more like having a game with the robot rather than helping it.} Furthermore, five participants highlighted the additional enjoyment that such games could bring to urban life, thus \emph{`adding colour to the city'} (p6). Another participant, p10, suggested that they would like to see more of these types of games, as they provide a good way \emph{`to kill the time [...] while waiting for the bus'}.

The most common negative feedback regarding the playful help-seeking experience centred on the extra effort required beyond simply assisting the robot (n=8). For example, in the traffic light scenario where participants were invited to press the button multiple times to complete a shooting game, four participants questioned the necessity of repeatedly pressing the button to complete the game. Additionally, five participants expressed a desire to skip the game and help the robot directly.

%[More play than help]
%[extra effort][option to skip the game]

\subsection{Attitudes towards the robot: acceptance, trust, and social attributes}

\begin{figure}[h]
\begin{center}
\includegraphics[width=0.5\textwidth]{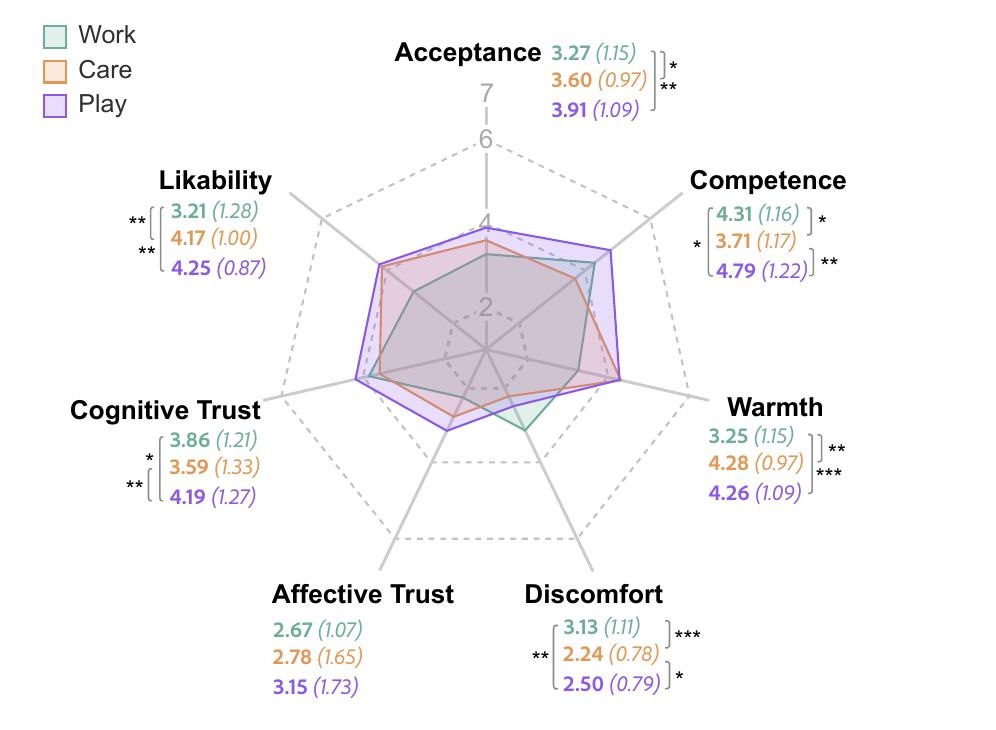}
\end{center}
\caption{Qualitative results of participants' attitudes towards the robot. *: p < .05, **: p < .01, ***: p < .001 }\label{impact}
\Description{Figure 7 shows a radar chart illustrating qualitative results of participants’ attitudes towards a robot, measured across seven attributes: Acceptance, Likability, Competence, Warmth, Cognitive Trust, Affective Trust, and Discomfort. Three strategies are evaluated: Work (gray), Care (green), and Play (purple). Each vertex of the radar represents one of the attributes with scores plotted and connected to form three distinct polygons. Notable scores include high Competence for Work at 4.31, high Warmth for Care at 4.26, and low Discomfort for Play at 2.24.}
\end{figure}
\subsubsection{Quantitative results}
Following descriptive data analysis (see Fig.~\ref{impact}), participants' acceptance of the robot was highest in \emph{Play} and lowest in \emph{Work}. Friedman's ANOVA showed significant differences ($\chi^2(3)$=9.156, p=0.01). Post-hoc tests indicated the acceptance scores in both \emph{Play} ($p=0.001$) and \emph{Care} \textbf{($p=0.04$)} were significantly higher than in \emph{Work}, with no significant differences between \emph{Care} and \emph{Play}.

Regarding the robot's social attributes, it received the lowest competence rating in \emph{Care} and the highest in \emph{Play}. The robot's warmth attributes received the highest rating in \emph{Care}, slightly lower in \emph{Play}, and the lowest in \emph{Work}. The robot received the lowest discomfort rating in \emph{Care} and the highest in \emph{Work}. Significant differences were found in Friedman's ANOVA test across all three dimensions (competence: $\chi^2(3)$=12.413, p=0.002; warmth: $\chi^2(3)$=13.130, p=0.001; discomfort: $\chi^2(3)$=13.505, p=0.001). Post-hoc tests indicated that the robot was rated significantly less competent in \emph{Care} compared to both \emph{Work} ($p=0.01$) and \emph{Play} ($p=0.001$), with \emph{Play} also scoring significantly higher than \emph{Work} ($p=0.01$). Warmth ratings were significantly higher in both \emph{Care} ($p=0.001$) and \emph{Play} ($p<0.001$) compared to \emph{Work}, with no significant differences between \emph{Care} ($p=0.03$) and \emph{Play}. Robots in both \emph{Care} ($p<0.001$) and \emph{Play} ($p=0.006$) were rated significantly lower for discomfort than those in \emph{Work}, and \emph{Care} also scoring significantly lower than \emph{Play} ($p=0.03$).   

Participants' trust was highest in \emph{Play} and lowest in \emph{Work} across both cognitive trust and affective trust subscales. \emph{Care} received the lowest cognitive trust rating, while \emph{Work} received the lowest affective trust rating. Friedman's ANOVA showed significant differences in cognitive trust ($\chi^2(3)$=7.670, p=0.022), but not in  affective trust ($\chi^2(3)$=5.365, p=0.068) subscales.
Post-hoc tests revealed that participants' cognitive trust ratings in \emph{Play} were significantly higher than in \emph{Work} ($p=0.01$) and \emph{Care} ($p=0.001$). There were no significant differences between \emph{Care} and \emph{Work} in the cognitive trust subscale.

At last, the robot in \emph{Work} was rated lowest in likability, followed by \emph{Care} and \emph{Play}. Friedman's ANOVA showed significant differences ($\chi^2(3)$=15.438, p<0.001). Post-hoc tests indicated the likability scores in both \emph{Play} ($p=0.001$) and \emph{Care} \textbf{($p=0.001$)} were significantly higher than in \emph{Work}, with no significant differences between \emph{Care} and \emph{Play}.

\subsubsection{Qualitative feedback on impressions of the robot}
The synthesised voice in \emph{Work} condition was perceived by six participants as machine-like or artificial. Three participants mentioned that the speech functionality aligned with their expectations of a robot, leading them to perceive the robot as \emph{`competitive'} (p14) and \emph{`professional'} (p17), while on the other hand three participants described the robot as \emph{`scary'} or \emph{`creepy'}. 

In the \emph{Care} condition, \emph{`cute'} was the impression most frequently attributed to the robot by participants (n=9). Additionally, we observed that two participants petted the robot upon their first encounter, as exemplified by p1, who expressed a desire to comfort it due to its sad facial expression. Many participants also indicated that the facial expression made the robot more human-like (n=7) or animal-like (n=6). However, the sad emotion also raised people's questions towards the robot's capability (n=6), making it less trustworthy (n=5) or even \emph{`stupid'} (n=2). P12 exemplified this sentiment by describing the robot with the sad face as \emph{`useless'}, likening its helplessness to the impression of \emph{`a baby'} and commenting, \emph{`It's like a baby, so I don't think I can trust the robot to do a lot of things.'} Furthermore, three participants felt manipulated by the robot displaying a sad face to prompt them to offer help. P3 noted, \emph{`The one with expressions feels like it’s a robot using tactics, like pretending to be pitiable.'}  Another intriguing response from p1 suggested they felt somewhat \emph{`being scammed'} for helping the robot.

In the \emph{Play} condition, five participants described the robot as \emph{`smart'} or \emph{`intelligent'}. Three participants indicated that they have more confidence in the robot's problem-solving abilities. P14 further elaborated on this point, indicating that the robot appeared \emph{`to have predicted that it would encounter such difficulties'} and was able to use a playful strategy \emph{`to overcome a particular scenario'}. Two participants described the robot as being \emph{`active'} (p16) and \emph{`sociable'} (p17), noting that it consistently sought the attention of passersby to engage them in game-playing. Furthermore, while four participants perceived the robot projecting a game onto the ground as less human-like, even comparing it to a \emph{`game console'} (P3), two other participants felt as though they were playing a game with a \emph{`kid'}.

\section{Discussion}
The section starts with three key considerations for facilitating robot help-seeking from bystanders generated from study results, then delves into a focused reflection on the game-inspired playful help-seeking strategy. 
%\subsection{Non-invasiveness over efficiency}
% Existing research focuses on apply Politeness Theory/ improve social framing of such request/ investigating the linguistic content of robot-issued requests, 

% ambiguity and politeness were found to be negatively correlated in previous studies~\cite{Budde2017Needy, Boos2022Delivery}, which is kind of supported by our results Playful is explicit, taking time to understand, but was good as they provide freedom for people to ignore

%\subsection{Balancing persuasiveness and efficiency}
%\subsection{Help-seeking characteristic (RQ1): balancing persuasiveness and efficiency}
%\subsection{Facilitating Bystander Helping Behaviours}

\subsection{Casting bystander help as voluntary engagement over obligated response}

Achieving task efficiency has long been one of the primary objectives in human-robot collaboration~\cite{ajoudani2018progressHRC}. Thus, verbal speech help-seeking has become the most common method for robots to express their need for assistance, due to its rich and intuitive communication, widely adopted in both research~\cite{Backhaus2018someone, Budde2017Needy, Cameron2015Button, knepper2015recovering, Srinivasan2016HelpMePlease} and commercially-deployed urban robots~\cite{dailymail_2023}. However, our study results raise questions about whether such explicit, direct, and objective-oriented approaches are suitable for situations where robots request help from bystanders. Despite its clarity and communication efficiency, verbal help-seeking was perceived negatively in terms of politeness and appropriateness, casting an unfavourable impression of the robot as also evidenced through the lowest levels of liking and acceptance among all tested conditions. Participants' comments revealed a conflict that emerged from their perceived role as passive bystanders without responsibility towards the robot, as opposed to the sense of command and obligation that they felt when receiving direct requests from the robot. This conflict in turn resulted in aversion and their reported reduced willingness to provide assistance.

In contrast, implicit methods such as emotional expression and playful engagement, while less efficient in terms of clear communication, were perceived as more polite and socially acceptable. For individuals who casually encounter urban robots in public spaces while being involved in a variety of activities and having different agendas, there is often a misalignment of goals with those of urban robots~\cite{pelikan2024encountering, Huttenrauch2003ToHelpOrNot}. Consequently, it becomes crucial that when robots seek assistance from these bystanders, they do so in a manner that minimises disruption to the activities these individuals are currently engaged in. Therefore, participants generally favoured the indirect and implicit nature of emotional expression or playful help-seeking, as these approaches engendered a feeling of reduced obligation, conveying to the bystander that they have the freedom to ignore the request and opt not to assist. Urban robot help-seeking strategies should therefore consider implicit and less persuasive approaches, presenting bystander assistance as a voluntary, spontaneous act driven by personal kindness or playful interest, rather than as an obligation, which is often associated with direct verbal requests.

% Help-seeking strategies such as playfulness and emotional expression, wh

%This finding indirectly contributes to the discourse on the relationship between ambiguity and politeness, a topic previously explored in verbal help-seeking contexts ~\cite{Budde2017Needy, Boos2022Delivery}. While our research does not investigate language-based help-seeking, participants' responses to the implicit and ambiguous ways of asking help (i.e., emotional expression and playful engagement) exhibit parallels to these established findings, especially in the conext of bystander .

\subsection{Incentivising bystander efforts through playful help-seeking strategy}
%\subsection{Incentivising  effort-intensive assistance through playful interatcion}
%\subsection{Incentivising bystander when effort required increase}
%%Impact of Request Intensity on Willingness to Assist and Mood Changes ... Playful
%%Incentivising bystander efforts through playful help-seeking strategy

Prior research suggests that more laborious requests from robots tend to reduce people's willingness to assist~\cite{Srinivasan2016HelpMePlease}. Our study explored this phenomenon across three distinct help-seeking scenarios, ranging from minimal (such as stepping aside to allow the robot to pass) to such that required more physical effort (like bending over to push a stuck robot). We observed a clear trend in \emph{Work} and \emph{Care} conditions, where participants demonstrated a reluctance to engage in the more physically demanding task of pushing the robot. Interestingly, participants' willingness to help in \emph{Play} conditions was consistently stable across all scenarios. Furthermore, the disparity between the playful strategy and the two other help-seeking strategies in terms of willingness to help was even more pronounced in the robot stuck scenario that had the highest request demand.
%and its improvement in willingness compared to the other two help-seeking strategies was more pronounced with larger request sizes in the robot stuck scenario.

This trend was also mirrored in participants' mood changes. In \emph{Work} and \emph{Care} conditions, more laborious requests (i.e., in the robot stuck scenario that required participants to bend over to push the robot) led to a decrease in their mood after helping the robot. Conversely, in \emph{Play} condition, the most effort-intensive robot stuck scenario resulted in the highest reported mood levels. Although various game-inspired concepts may have influenced the outcomes, participants primarily attributed the elevated mood levels in the effort-intensive scenarios to an increased sense of achievement and competence, which is the key motivational impact of gaming elements~\cite{Tondello2016}. This experience effectively neutralised the perceived effort and the accompanying sense of fatigue. 

% While acknowledging that the variety of game-inspired concepts might have played a role, participants primarily attributed the elevated mood levels in the more effort-intensive scenario to a heightened sense of achievement and competence, the key motivational impact of gaming elements~\cite{Tondello2016}. 

Drawing upon the principles of reciprocity and altruism in social exchange theory ~\cite{Russell2005SocialExchange}, our study highlights that bystanders, who do not directly benefit from urban robot services, often need a form of incentives for their time and effort in assisting robots they encounter casually. Such needs are particularly significant when the physical effort from helpers increases, highlighting the lack of direct mutual benefits. While playful help-seeking strategy, leveraging gameful experiences, offers a potential solution, using game elements as motivators could also raise ethical considerations about invisible labour~\cite{dubbell2015invisible} – a concern also noted in socio-technical research on bystander assistance for commercially deployed robots~\cite{hakli2023helping}. While participants in our study generally expressed autonomy in deciding whether to engage in gameplay, it is important to be mindful of the potential for creating invisible labour when designing gameful incentives.

\subsection{Avoiding helpless robot portrayal}

Public trust and acceptance are crucial for the successful integration of robots into urban environments. There have been instances where distrust by the general public even led to regulatory bans on urban robots~\cite{machines2018Regulatory}. Despite this, the specific impact of robot help-seeking behaviour on people's additutdes towards robots has not been explored in existing research.

Previous research has shown that robots displaying sad emotions are more likely to receive human assistance, both in collaborative settings~\cite{Zhou2020Sad,urakami2023emotional} and when used in robot pets seeking help~\cite{Daly2020RobotInNeed}. Our research expands upon the investigation of using emotional expressions to elicit help from bystanders, uncovering patterns that diverge from those in more traditional human-robot collaboration settings. While emotional expressions are generally effective in eliciting help in human-robot teaming~\cite{urakami2023emotional}, our findings suggest that in casual bystander contexts, a robot displaying sadness was perceived as needy and less competent, which at the same time resulted in a decrease in people's cognitive trust. These expressions even further engendered a perception of emotional blackmail, especially without an established human-robot relationship between bystanders. This aligns with prior findings~\cite{Yu2024HelpSeeking} that bystanders tend to view urban service robots as professional entities, thereby expecting them to exhibit proficiency and dependability.
 
In contrast, robots employing playful strategies were perceived as more capable, as they were perceived as actively helping themselves out of situations using well-prepared and dedicated strategies. This approach further resulted in enhanced trust and acceptance towards the robot. We therefore conclude that urban robot help-seeking should avoid portraying robots as helpless or needy in order to minimise the negative impact on bystander trust in the robot.

\subsection{Reflections on the game-inspired playful help-seeking design concepts}

%--Intuitiveness
Our user study demonstrated that integrating mechanics and visual cues from classic games has the potential to effectively engage bystanders in playful human-robot collaborations as a means to provide assistance to a robot. Notably, this was also true for the two participants (p7 and p13) from the older demographic group. Analysis of their willingness to help results in \emph{Play} condition revealed that neither participant was statistically deviant from the overall sample, with standard deviations of 1.352 and 0.0854, respectively.
Particularly, p7, who is over 65 and was not familiar with some of the games but managed to comprehend and engage successfully. However, while most participants could recognise and comprehend the game's functions, there was some ambiguity about whether \emph{they} were invited by the robot to participate. Noteworthy, this issue did not arise in the Tetris-inspired concept, as the movement of the tetra block in response to the participants' actions made it immediately clear that they had full control over the game elements. Conversely, in the other scenarios requiring mediation of other objects for game element manipulation~(i.e.~pressing the traffic light button to shoot the projectile or pushing the robot to chase the ghosts), participants faced a steeper learning curve and expressed uncertainty regarding their involvement. These findings suggest that when designing game-inspired playful help-seeking requests, game elements should be designed in a way that they respond instantly to bystander movement or incorporate a feed-forward mechanism~\cite{Nguyen2023}.  
%direct manipulation of game elements that respond instantly to bystander actions should be considered.

Repetition is a fundamental aspect of game design, serving as a key reinforcement of game mechanics and thereby enhancing player engagement ~\cite{ROBSON2015411GamePrinciples}. Our design concepts also employed repetitive elements; for instance, bystanders were required to press the traffic light button multiple times to eliminate all enemies. 
While enjoyable for some participants, others noted a disconnect between the game context, requiring repetitive action, and the real-life context, where a single press of the traffic light button is typically sufficient. This feedback highlights a potential mismatch between game mechanics and real-world expectations. Furthermore, several participants indicated a preference for omitting repetitive game-play actions if they were in a rush. Considering the rapid pace of urban life, it is evident that designing for game-inspired playful help-seeking must be succinct and allow for flexibility in engagement, accommodating the varying preferences and time constraints of bystanders. %These could be achieved by more responsive interactions through the advanced robot perception system to recognise people's intentions. 
%[--- impact on the urban environment ]

Moreover, it's crucial to consider the broader urban environment where such playful engagements for robot help-seeking are situated. The novelty effect observed in our study, if implemented in the real world, necessitates consideration of the potential impact of people crowding, as people may gather around the interaction area. Furthermore, since the game-playing interaction is attention-demanding, it could distract bystanders from cautiously observing their surrounding traffic environment, potentially leading to safety concerns. Additionally, the urban environment's rhythm, busy on workdays and relaxed on weekends, could affect bystander engagement levels. Game-inspired playful help-seeking design should thus adapt to these varying urban rhythms to better align with people's availability and willingness to engage.

% Prior research indicates that larger robot help requests decrease people's willingness to offer assistance ~\cite{Srinivasan2016HelpMePlease}.
% Our study also assessed help-seeking across three scenarios, varying the robot's help-seeking request size from small (stepping aside to make room for the robot to pass) to larger ones that required participants to bend over and push the robot out of being stuck. 

% participants liked the faulty robot significantly better than the robot that interacted flawlessly /  results showed that the participants liked the faulty robot significantly more than the flawless one~\cite{mirnig2017err}.

% more willing if people perceive robot as their collabrator,\cite{Zhou2020Sad,urakami2023emotional} which is not the same case for bystanders and urban robot.  Different from our study results/ could negatively impact people's trust if using strategies like emotional help-seekings. 

% \subsection{Considering potential impacts on urban environment}

% noise/ safety/ extra benefits/ 
% Contextual adaptability issues brought up in OzCHI workshop.

\subsection{Limitations}

First, the controlled lab study design may not fully reflect spontaneous real-life helping behaviour, as some participants noted potential differences in their decisions. In addition, the scenarios included in our study could be subject to the potential bias of social media posts, as they are derived from a previous online ethnography study. Conducting field studies with real robots and naive bystanders across more diverse scenarios could further validate our findings. 
Second, the ecological validity of this work is limited by the use of a VR simulation. The novelty of the VR experience influenced participants' mood assessments, potentially leading to inflated mood ratings after the VR familiarisation session. This may explain why the `helper's high' phenomenon -- helping someone or something else can lead to psychological benefits -- was not validated in our study. Furthermore, implementing the playful design concepts in VR overlooks potential technical challenges in real-world implementations. For instance, limitations exist in projecting in outdoor environments due to variations in lighting and distance constraints. While the VR study validates the playful help-seeking strategy as a proof of concept, further technological refinement and adaptation are required for practical real-world application. 
Third, our study findings are based on a small group of participants with Western cultural or educational backgrounds, which might have facilitated their understanding of such game-playing interactions. To ensure the generalisability of these findings, further validation is necessary with participants from a more diverse range of cultural backgrounds.

% \subsection{Impacts} 

% [achieving efficiency and effectiveness is not the only goal]
% [non-invasiveness or effciency]

% participants liked the faulty robot significantly better than the robot that interacted flawlessly /  results showed that the participants liked the faulty robot significantly more than the flawless one~\cite{mirnig2017err}.

% [Differences in Relationship ]participant collaborate with robot in a lego building 

% warmth associated with humanoids can be beneficial during the service recovery process. ~\cite{Choi2021Humanoid}

% [using emotions] more willing if people perceive robot as their collabrator,\cite{Zhou2020Sad,urakami2023emotional} which is not the same case for bystanders and urban robot

% [consider impacts on the urban contexts] 
% noise/ safety/ extra benefits/ 

\section{Conclusion}
Our study extended the use case of playful engagement in HRI from creating enjoyable experiences to facilitating casual collaboration. The evaluation study in VR with 24 participants found that playful engagement has the potential to enhance bystanders' willingness to assist and improve their mood after helping the robot. In our study, this effect was particularly noticeable when the assistance required more effort. By comparing the helping-as-play strategy with helping-as-work (using verbal help-seeking) and helping-as-care (using emotional expression) strategies, we found that participants perceived verbal help-seeking to be impolite and that there was a decline in trust towards the robot when using sad emotions to request help. These findings prompt broader considerations for robot help-seeking in urban contexts and offer insights into playful help-seeking as a strategy to obtain assistance from casual bystanders in dynamic and unpredictable urban environments.
%design that can inform future explorations in leveraging playful engagement for casual human-robot collaboration.

\begin{acks}
This study is funded by the Australian Research Council through the ARC Discovery Project DP220102019 Shared-Space Interactions Between People and Autonomous Vehicles. We thank all the participants for taking part in this research. We also thank the anonymous DIS'24 reviewers and ACs for their constructive feedback and suggestions to make this contribution stronger. 
\end{acks}
\balance{}
\bibliographystyle{ACM-Reference-Format}
\bibliography{sample-base}


\begin{thebibliography}{96}

%%% the Web address.

\ifx \showCODEN    \undefined \def \showCODEN     #1{\unskip}     \fi
\ifx \showDOI      \undefined \def \showDOI       #1{#1}\fi
\ifx \showISBNx    \undefined \def \showISBNx     #1{\unskip}     \fi
\ifx \showISBNxiii \undefined \def \showISBNxiii  #1{\unskip}     \fi
\ifx \showISSN     \undefined \def \showISSN      #1{\unskip}     \fi
\ifx \showLCCN     \undefined \def \showLCCN      #1{\unskip}     \fi
\ifx \shownote     \undefined \def \shownote      #1{#1}          \fi
\ifx \showarticletitle \undefined \def \showarticletitle #1{#1}   \fi
\ifx \showURL      \undefined \def \showURL       {\relax}        \fi
% invisible to TeX
\providecommand\bibfield[2]{#2}
\providecommand\bibinfo[2]{#2}
\providecommand\natexlab[1]{#1}
\providecommand\showeprint[2][]{arXiv:#2}

\bibitem[Ajoudani et~al\mbox{.}(2018)]%
        {ajoudani2018progressHRC}
\bibfield{author}{\bibinfo{person}{Arash Ajoudani}, \bibinfo{person}{Andrea~Maria Zanchettin}, \bibinfo{person}{Serena Ivaldi}, \bibinfo{person}{Alin Albu-Sch{\"a}ffer}, \bibinfo{person}{Kazuhiro Kosuge}, {and} \bibinfo{person}{Oussama Khatib}.} \bibinfo{year}{2018}\natexlab{}.
\newblock \showarticletitle{Progress and prospects of the human--robot collaboration}.
\newblock \bibinfo{journal}{\emph{Autonomous Robots}}  \bibinfo{volume}{42} (\bibinfo{year}{2018}), \bibinfo{pages}{957--975}.
\newblock
\urldef\tempurl%
\url{https://doi.org/10.1007/s10514-017-9677-2}
\showDOI{\tempurl}


\bibitem[Babel et~al\mbox{.}(2022)]%
        {Babel2022findings}
\bibfield{author}{\bibinfo{person}{Franziska Babel}, \bibinfo{person}{Johannes Kraus}, {and} \bibinfo{person}{Martin Baumann}.} \bibinfo{year}{2022}\natexlab{}.
\newblock \showarticletitle{Findings From A Qualitative Field Study with An Autonomous Robot in Public: Exploration of User Reactions and Conflicts}.
\newblock \bibinfo{journal}{\emph{International Journal of Social Robotics}} \bibinfo{volume}{14}, \bibinfo{number}{7} (\bibinfo{year}{2022}), \bibinfo{pages}{1625--1655}.
\newblock


\bibitem[Backhaus et~al\mbox{.}(2018)]%
        {Backhaus2018someone}
\bibfield{author}{\bibinfo{person}{Nils Backhaus}, \bibinfo{person}{Patricia~H. Rosen}, \bibinfo{person}{Andrea Scheidig}, \bibinfo{person}{Horst-Michael Gross}, {and} \bibinfo{person}{Sascha Wischniewski}.} \bibinfo{year}{2018}\natexlab{}.
\newblock \showarticletitle{Somebody help me, please?!” Interaction Design Framework for Needy Mobile Service Robots}. In \bibinfo{booktitle}{\emph{2018 IEEE Workshop on Advanced Robotics and its Social Impacts (ARSO)}}. \bibinfo{publisher}{IEEE}, \bibinfo{address}{Genova, Italy}, \bibinfo{pages}{54--61}.
\newblock
\urldef\tempurl%
\url{https://doi.org/10.1109/ARSO.2018.8625721}
\showDOI{\tempurl}


\bibitem[Bartneck et~al\mbox{.}(2009)]%
        {bartneck2009GodSpeed}
\bibfield{author}{\bibinfo{person}{Christoph Bartneck}, \bibinfo{person}{Dana Kuli{\'c}}, \bibinfo{person}{Elizabeth Croft}, {and} \bibinfo{person}{Susana Zoghbi}.} \bibinfo{year}{2009}\natexlab{}.
\newblock \showarticletitle{Measurement instruments for the anthropomorphism, animacy, likeability, perceived intelligence, and perceived safety of robots}.
\newblock \bibinfo{journal}{\emph{International journal of social robotics}}  \bibinfo{volume}{1} (\bibinfo{year}{2009}), \bibinfo{pages}{71--81}.
\newblock
\urldef\tempurl%
\url{https://doi.org/10.1007/s12369-008-0001-3}
\showDOI{\tempurl}


\bibitem[Boos et~al\mbox{.}(2022)]%
        {Boos2022Delivery}
\bibfield{author}{\bibinfo{person}{Annika Boos}, \bibinfo{person}{Markus Zimmermann}, \bibinfo{person}{Monika Zych}, {and} \bibinfo{person}{Klaus Bengler}.} \bibinfo{year}{2022}\natexlab{}.
\newblock \showarticletitle{Polite and Unambiguous Requests Facilitate Willingness to Help an Autonomous Delivery Robot and Favourable Social Attributions}. In \bibinfo{booktitle}{\emph{2022 31st IEEE International Conference on Robot and Human Interactive Communication (RO-MAN)}}. \bibinfo{publisher}{IEEE}, \bibinfo{address}{Naples, Italy}, \bibinfo{pages}{1620--1626}.
\newblock
\urldef\tempurl%
\url{https://doi.org/10.1109/RO-MAN53752.2022.9900870}
\showDOI{\tempurl}


\bibitem[Booth et~al\mbox{.}(2017)]%
        {Piggybacking2017Booth}
\bibfield{author}{\bibinfo{person}{Serena Booth}, \bibinfo{person}{James Tompkin}, \bibinfo{person}{Hanspeter Pfister}, \bibinfo{person}{Jim Waldo}, \bibinfo{person}{Krzysztof Gajos}, {and} \bibinfo{person}{Radhika Nagpal}.} \bibinfo{year}{2017}\natexlab{}.
\newblock \showarticletitle{Piggybacking Robots: Human-Robot Overtrust in University Dormitory Security}. In \bibinfo{booktitle}{\emph{Proceedings of the 2017 ACM/IEEE International Conference on Human-Robot Interaction}} (Vienna, Austria) \emph{(\bibinfo{series}{HRI '17})}. \bibinfo{publisher}{Association for Computing Machinery}, \bibinfo{address}{New York, NY, USA}, \bibinfo{pages}{426–434}.
\newblock
\showISBNx{9781450343367}
\urldef\tempurl%
\url{https://doi.org/10.1145/2909824.3020211}
\showDOI{\tempurl}


\bibitem[Bradley(1958)]%
        {bradley1958complete}
\bibfield{author}{\bibinfo{person}{James~V Bradley}.} \bibinfo{year}{1958}\natexlab{}.
\newblock \showarticletitle{Complete counterbalancing of immediate sequential effects in a Latin square design}.
\newblock \bibinfo{journal}{\emph{J. Amer. Statist. Assoc.}} \bibinfo{volume}{53}, \bibinfo{number}{282} (\bibinfo{year}{1958}), \bibinfo{pages}{525--528}.
\newblock


\bibitem[Braun and Clarke(2006)]%
        {braun2006thematic}
\bibfield{author}{\bibinfo{person}{Virginia Braun} {and} \bibinfo{person}{Victoria Clarke}.} \bibinfo{year}{2006}\natexlab{}.
\newblock \showarticletitle{Using thematic analysis in psychology}.
\newblock \bibinfo{journal}{\emph{Qualitative Research in Psychology}} \bibinfo{volume}{3}, \bibinfo{number}{2} (\bibinfo{year}{2006}), \bibinfo{pages}{77--101}.
\newblock
\urldef\tempurl%
\url{https://doi.org/10.1191/1478088706qp063oa}
\showDOI{\tempurl}
\showeprint{https://www.tandfonline.com/doi/pdf/10.1191/1478088706qp063oa}


\bibitem[Br\v{s}\v{c}i\'{c} et~al\mbox{.}(2015)]%
        {ChildrenAbuse2015Brvsvcic}
\bibfield{author}{\bibinfo{person}{Dra\v{z}en Br\v{s}\v{c}i\'{c}}, \bibinfo{person}{Hiroyuki Kidokoro}, \bibinfo{person}{Yoshitaka Suehiro}, {and} \bibinfo{person}{Takayuki Kanda}.} \bibinfo{year}{2015}\natexlab{}.
\newblock \showarticletitle{Escaping from Children's Abuse of Social Robots}. In \bibinfo{booktitle}{\emph{Proceedings of the Tenth Annual ACM/IEEE International Conference on Human-Robot Interaction}} (Portland, Oregon, USA) \emph{(\bibinfo{series}{HRI '15})}. \bibinfo{publisher}{Association for Computing Machinery}, \bibinfo{address}{New York, NY, USA}, \bibinfo{pages}{59–66}.
\newblock
\showISBNx{9781450328838}
\urldef\tempurl%
\url{https://doi.org/10.1145/2696454.2696468}
\showDOI{\tempurl}


\bibitem[Bu et~al\mbox{.}(2023)]%
        {TrashBarrel2023Bu}
\bibfield{author}{\bibinfo{person}{Fanjun Bu}, \bibinfo{person}{Ilan Mandel}, \bibinfo{person}{Wen-Ying Lee}, {and} \bibinfo{person}{Wendy Ju}.} \bibinfo{year}{2023}\natexlab{}.
\newblock \showarticletitle{Trash Barrel Robots in the City}. In \bibinfo{booktitle}{\emph{Companion of the 2023 ACM/IEEE International Conference on Human-Robot Interaction}} (<conf-loc>, <city>Stockholm</city>, <country>Sweden</country>, </conf-loc>) \emph{(\bibinfo{series}{HRI '23})}. \bibinfo{publisher}{Association for Computing Machinery}, \bibinfo{address}{New York, NY, USA}, \bibinfo{pages}{875–877}.
\newblock
\showISBNx{9781450399708}
\urldef\tempurl%
\url{https://doi.org/10.1145/3568294.3580206}
\showDOI{\tempurl}


\bibitem[Budde et~al\mbox{.}(2018)]%
        {Budde2017Needy}
\bibfield{author}{\bibinfo{person}{Vanessa Budde}, \bibinfo{person}{Nils Backhaus}, \bibinfo{person}{Patricia~H. Rosen}, {and} \bibinfo{person}{Sascha Wischniewski}.} \bibinfo{year}{2018}\natexlab{}.
\newblock \showarticletitle{Needy Robots - Designing Requests for Help Using Insights from Social Psychology}. In \bibinfo{booktitle}{\emph{2018 IEEE Workshop on Advanced Robotics and Its Social Impacts (ARSO)}}. \bibinfo{publisher}{IEEE Press}, \bibinfo{address}{Genova, Italy}, \bibinfo{pages}{48–53}.
\newblock
\urldef\tempurl%
\url{https://doi.org/10.1109/ARSO.2018.8625724}
\showDOI{\tempurl}


\bibitem[Cameron et~al\mbox{.}(2015)]%
        {Cameron2015Button}
\bibfield{author}{\bibinfo{person}{David Cameron}, \bibinfo{person}{Emily~C. Collins}, \bibinfo{person}{Adriel Chua}, \bibinfo{person}{Samuel Fernando}, \bibinfo{person}{Owen McAree}, \bibinfo{person}{Uriel Martinez-Hernandez}, \bibinfo{person}{Jonathan~M. Aitken}, \bibinfo{person}{Luke Boorman}, {and} \bibinfo{person}{James Law}.} \bibinfo{year}{2015}\natexlab{}.
\newblock \showarticletitle{Help! I Can't Reach the Buttons: Facilitating Helping Behaviors Towards Robots}. In \bibinfo{booktitle}{\emph{Biomimetic and Biohybrid Systems}}, \bibfield{editor}{\bibinfo{person}{Stuart~P. Wilson}, \bibinfo{person}{Paul~F.M.J. Verschure}, \bibinfo{person}{Anna Mura}, {and} \bibinfo{person}{Tony~J. Prescott}} (Eds.). \bibinfo{publisher}{Springer International Publishing}, \bibinfo{address}{Cham}, \bibinfo{pages}{354--358}.
\newblock
\showISBNx{978-3-319-22979-9}


\bibitem[Carpinella et~al\mbox{.}(2017)]%
        {Carpinella2017RoSAS}
\bibfield{author}{\bibinfo{person}{Colleen~M. Carpinella}, \bibinfo{person}{Alisa~B. Wyman}, \bibinfo{person}{Michael~A. Perez}, {and} \bibinfo{person}{Steven~J. Stroessner}.} \bibinfo{year}{2017}\natexlab{}.
\newblock \showarticletitle{The Robotic Social Attributes Scale (RoSAS): Development and Validation}. In \bibinfo{booktitle}{\emph{Proceedings of the 2017 ACM/IEEE International Conference on Human-Robot Interaction}} (Vienna, Austria) \emph{(\bibinfo{series}{HRI '17})}. \bibinfo{publisher}{Association for Computing Machinery}, \bibinfo{address}{New York, NY, USA}, \bibinfo{pages}{254–262}.
\newblock
\showISBNx{9781450343367}
\urldef\tempurl%
\url{https://doi.org/10.1145/2909824.3020208}
\showDOI{\tempurl}


\bibitem[Castellano et~al\mbox{.}(2019)]%
        {Castellano2019Waste}
\bibfield{author}{\bibinfo{person}{Giovanna Castellano}, \bibinfo{person}{Berardina~De Carolis}, \bibinfo{person}{Nicola Macchiarulo}, {and} \bibinfo{person}{Veronica Rossano}.} \bibinfo{year}{2019}\natexlab{}.
\newblock \showarticletitle{Learning waste Recycling by playing with a Social Robot}. In \bibinfo{booktitle}{\emph{2019 IEEE International Conference on Systems, Man and Cybernetics (SMC)}}. \bibinfo{publisher}{IEEE}, \bibinfo{address}{Bari, Italy}, \bibinfo{pages}{3805--3810}.
\newblock
\urldef\tempurl%
\url{https://doi.org/10.1109/SMC.2019.8914455}
\showDOI{\tempurl}


\bibitem[Cha and Matarić(2016)]%
        {Cha2016nonverbal}
\bibfield{author}{\bibinfo{person}{Elizabeth Cha} {and} \bibinfo{person}{Maja Matarić}.} \bibinfo{year}{2016}\natexlab{}.
\newblock \showarticletitle{Using nonverbal signals to request help during human-robot collaboration}. In \bibinfo{booktitle}{\emph{2016 IEEE/RSJ International Conference on Intelligent Robots and Systems (IROS)}}. \bibinfo{publisher}{IEEE}, \bibinfo{address}{Daejeon, Korea}, \bibinfo{pages}{5070--5076}.
\newblock
\urldef\tempurl%
\url{https://doi.org/10.1109/IROS.2016.7759744}
\showDOI{\tempurl}


\bibitem[Chen et~al\mbox{.}(2023)]%
        {chen2023gamified}
\bibfield{author}{\bibinfo{person}{Tan-I Chen}, \bibinfo{person}{Shih-Kai Lin}, {and} \bibinfo{person}{Hung-Chang Chung}.} \bibinfo{year}{2023}\natexlab{}.
\newblock \showarticletitle{Gamified Educational Robots Lead an Increase in Motivation and Creativity in STEM Education.}
\newblock \bibinfo{journal}{\emph{Journal of Baltic Science Education}} \bibinfo{volume}{22}, \bibinfo{number}{3} (\bibinfo{year}{2023}), \bibinfo{pages}{427--438}.
\newblock


\bibitem[Chirapornchai et~al\mbox{.}(2021)]%
        {Chirapornchai2021HelperSHigh}
\bibfield{author}{\bibinfo{person}{Chatchai Chirapornchai}, \bibinfo{person}{Paul Bremner}, {and} \bibinfo{person}{Joseph~E. Daly}.} \bibinfo{year}{2021}\natexlab{}.
\newblock \showarticletitle{Helper's High with a Robot Pet}. In \bibinfo{booktitle}{\emph{Companion of the 2021 ACM/IEEE International Conference on Human-Robot Interaction}} (Boulder, CO, USA) \emph{(\bibinfo{series}{HRI '21 Companion})}. \bibinfo{publisher}{Association for Computing Machinery}, \bibinfo{address}{New York, NY, USA}, \bibinfo{pages}{229–233}.
\newblock
\showISBNx{9781450382908}
\urldef\tempurl%
\url{https://doi.org/10.1145/3434074.3447165}
\showDOI{\tempurl}


\bibitem[Chowdhury et~al\mbox{.}(2021)]%
        {Chowdhury2021PlayHRC}
\bibfield{author}{\bibinfo{person}{Aparajita Chowdhury}, \bibinfo{person}{Aino Ahtinen}, \bibinfo{person}{Roel Pieters}, {and} \bibinfo{person}{Kaisa Väänänen}.} \bibinfo{year}{2021}\natexlab{}.
\newblock \showarticletitle{"How are you today, Panda the Robot?" – Affectiveness, Playfulness and Relatedness in Human-Robot Collaboration in the Factory Context}. In \bibinfo{booktitle}{\emph{2021 30th IEEE International Conference on Robot \& Human Interactive Communication (RO-MAN)}}. \bibinfo{publisher}{IEEE}, \bibinfo{address}{Vancouver, Canada}, \bibinfo{pages}{1089--1096}.
\newblock
\urldef\tempurl%
\url{https://doi.org/10.1109/RO-MAN50785.2021.9515351}
\showDOI{\tempurl}


\bibitem[Cila(2022)]%
        {Cila2022HAC}
\bibfield{author}{\bibinfo{person}{Nazli Cila}.} \bibinfo{year}{2022}\natexlab{}.
\newblock \showarticletitle{Designing Human-Agent Collaborations: Commitment, Responsiveness, and Support}. In \bibinfo{booktitle}{\emph{Proceedings of the 2022 CHI Conference on Human Factors in Computing Systems}} (New Orleans, LA, USA) \emph{(\bibinfo{series}{CHI '22})}. \bibinfo{publisher}{Association for Computing Machinery}, \bibinfo{address}{New York, NY, USA}, Article \bibinfo{articleno}{420}, \bibinfo{numpages}{18}~pages.
\newblock
\showISBNx{9781450391573}
\urldef\tempurl%
\url{https://doi.org/10.1145/3491102.3517500}
\showDOI{\tempurl}


\bibitem[Connolly et~al\mbox{.}(2020)]%
        {prosocial2020Connolly}
\bibfield{author}{\bibinfo{person}{Joe Connolly}, \bibinfo{person}{Viola Mocz}, \bibinfo{person}{Nicole Salomons}, \bibinfo{person}{Joseph Valdez}, \bibinfo{person}{Nathan Tsoi}, \bibinfo{person}{Brian Scassellati}, {and} \bibinfo{person}{Marynel V\'{a}zquez}.} \bibinfo{year}{2020}\natexlab{}.
\newblock \showarticletitle{Prompting Prosocial Human Interventions in Response to Robot Mistreatment}. In \bibinfo{booktitle}{\emph{Proceedings of the 2020 ACM/IEEE International Conference on Human-Robot Interaction}} (Cambridge, United Kingdom) \emph{(\bibinfo{series}{HRI '20})}. \bibinfo{publisher}{Association for Computing Machinery}, \bibinfo{address}{New York, NY, USA}, \bibinfo{pages}{211–220}.
\newblock
\showISBNx{9781450367462}
\urldef\tempurl%
\url{https://doi.org/10.1145/3319502.3374781}
\showDOI{\tempurl}


\bibitem[Creswell(2014)]%
        {Creswell2014MixMethod}
\bibfield{author}{\bibinfo{person}{John~W. Creswell}.} \bibinfo{year}{2014}\natexlab{}.
\newblock \bibinfo{booktitle}{\emph{A concise introduction to mixed methods research}}.
\newblock \bibinfo{publisher}{Sage}, \bibinfo{address}{Thousand Oaks, California, USA}.
\newblock


\bibitem[Cropanzano and Mitchell(2005)]%
        {Russell2005SocialExchange}
\bibfield{author}{\bibinfo{person}{Russell Cropanzano} {and} \bibinfo{person}{Marie~S. Mitchell}.} \bibinfo{year}{2005}\natexlab{}.
\newblock \showarticletitle{Social Exchange Theory: An Interdisciplinary Review}.
\newblock \bibinfo{journal}{\emph{Journal of Management}} \bibinfo{volume}{31}, \bibinfo{number}{6} (\bibinfo{year}{2005}), \bibinfo{pages}{874--900}.
\newblock
\urldef\tempurl%
\url{https://doi.org/10.1177/0149206305279602}
\showDOI{\tempurl}
\showeprint{https://doi.org/10.1177/0149206305279602}


\bibitem[Daly et~al\mbox{.}(2020)]%
        {Daly2020RobotInNeed}
\bibfield{author}{\bibinfo{person}{Joseph Daly}, \bibinfo{person}{Ute Leonards}, {and} \bibinfo{person}{Paul Bremner}.} \bibinfo{year}{2020}\natexlab{}.
\newblock \showarticletitle{Robots in Need: How Patterns of Emotional Behavior Influence Willingness to Help}. In \bibinfo{booktitle}{\emph{Companion of the 2020 ACM/IEEE International Conference on Human-Robot Interaction}} (Cambridge, United Kingdom) \emph{(\bibinfo{series}{HRI '20})}. \bibinfo{publisher}{Association for Computing Machinery}, \bibinfo{address}{New York, NY, USA}, \bibinfo{pages}{174–176}.
\newblock
\showISBNx{9781450370578}
\urldef\tempurl%
\url{https://doi.org/10.1145/3371382.3378301}
\showDOI{\tempurl}


\bibitem[Dobrosovestnova et~al\mbox{.}(2022)]%
        {Dobrosovestnova2022WithLittleHelp}
\bibfield{author}{\bibinfo{person}{Anna Dobrosovestnova}, \bibinfo{person}{Isabel Schwaninger}, {and} \bibinfo{person}{Astrid Weiss}.} \bibinfo{year}{2022}\natexlab{}.
\newblock \showarticletitle{With a Little Help of Humans. An Exploratory Study of Delivery Robots Stuck in Snow}. In \bibinfo{booktitle}{\emph{2022 31st IEEE International Conference on Robot and Human Interactive Communication (RO-MAN)}}. \bibinfo{publisher}{IEEE Press}, \bibinfo{address}{Napoli, Italy}, \bibinfo{pages}{1023–1029}.
\newblock
\urldef\tempurl%
\url{https://doi.org/10.1109/RO-MAN53752.2022.9900588}
\showDOI{\tempurl}


\bibitem[Donnermann et~al\mbox{.}(2021)]%
        {Donnermann2021}
\bibfield{author}{\bibinfo{person}{Melissa Donnermann}, \bibinfo{person}{Martina Lein}, \bibinfo{person}{Tanja Messingschlager}, \bibinfo{person}{Anna Riedmann}, \bibinfo{person}{Philipp Schaper}, \bibinfo{person}{Sophia Steinhaeusser}, {and} \bibinfo{person}{Birgit Lugrin}.} \bibinfo{year}{2021}\natexlab{}.
\newblock \showarticletitle{Social Robots and Gamification for technology supported learning: An empirical study on engagement and motivation}.
\newblock \bibinfo{journal}{\emph{Computers in Human Behavior}}  \bibinfo{volume}{121} (\bibinfo{date}{03} \bibinfo{year}{2021}), \bibinfo{pages}{106792}.
\newblock
\urldef\tempurl%
\url{https://doi.org/10.1016/j.chb.2021.106792}
\showDOI{\tempurl}


\bibitem[Dubbell(2015)]%
        {dubbell2015invisible}
\bibfield{author}{\bibinfo{person}{Julian Dubbell}.} \bibinfo{year}{2015}\natexlab{}.
\newblock \showarticletitle{Invisible labor, invisible play: Online gold farming and the boundary between jobs and games}.
\newblock \bibinfo{journal}{\emph{Vand. J. Ent. \& Tech. L.}}  \bibinfo{volume}{18} (\bibinfo{year}{2015}), \bibinfo{pages}{419}.
\newblock


\bibitem[Feingold-Polak et~al\mbox{.}(2021)]%
        {Feingold-Polak2021}
\bibfield{author}{\bibinfo{person}{Ronit Feingold-Polak}, \bibinfo{person}{Oren Barzel}, {and} \bibinfo{person}{Shelly Levy-Tzedek}.} \bibinfo{year}{2021}\natexlab{}.
\newblock \showarticletitle{A robot goes to rehab: a novel gamified system for long-term stroke rehabilitation using a socially assistive robot---methodology and usability testing}.
\newblock \bibinfo{journal}{\emph{Journal of NeuroEngineering and Rehabilitation}} \bibinfo{volume}{18}, \bibinfo{number}{1} (\bibinfo{date}{28 Jul} \bibinfo{year}{2021}), \bibinfo{pages}{122}.
\newblock
\showISSN{1743-0003}
\urldef\tempurl%
\url{https://doi.org/10.1186/s12984-021-00915-2}
\showDOI{\tempurl}


\bibitem[Fischer et~al\mbox{.}(2014)]%
        {Fischer2014SocialFrame}
\bibfield{author}{\bibinfo{person}{K. Fischer}, \bibinfo{person}{B. Soto}, \bibinfo{person}{C. Pantofaru}, {and} \bibinfo{person}{L. Takayama}.} \bibinfo{year}{2014}\natexlab{}.
\newblock \showarticletitle{Initiating interactions in order to get help: Effects of social framing on people's responses to robots' requests for assistance}. In \bibinfo{booktitle}{\emph{The 23rd IEEE International Symposium on Robot and Human Interactive Communication}}. \bibinfo{publisher}{IEEE}, \bibinfo{address}{Edinburgh, Scotland}, \bibinfo{pages}{999--1005}.
\newblock
\urldef\tempurl%
\url{https://doi.org/10.1109/ROMAN.2014.6926383}
\showDOI{\tempurl}


\bibitem[Gehrke et~al\mbox{.}(2023)]%
        {GEHRKE2023Observing}
\bibfield{author}{\bibinfo{person}{Steven~R. Gehrke}, \bibinfo{person}{Christopher~D. Phair}, \bibinfo{person}{Brendan~J. Russo}, {and} \bibinfo{person}{Edward~J. Smaglik}.} \bibinfo{year}{2023}\natexlab{}.
\newblock \showarticletitle{Observed sidewalk autonomous delivery robot interactions with pedestrians and bicyclists}.
\newblock \bibinfo{journal}{\emph{Transportation Research Interdisciplinary Perspectives}}  \bibinfo{volume}{18} (\bibinfo{year}{2023}), \bibinfo{pages}{100789}.
\newblock
\showISSN{2590-1982}
\urldef\tempurl%
\url{https://doi.org/10.1016/j.trip.2023.100789}
\showDOI{\tempurl}


\bibitem[Grzeskowiak et~al\mbox{.}(2020)]%
        {VR2020Grzeskowiak}
\bibfield{author}{\bibinfo{person}{Fabien Grzeskowiak}, \bibinfo{person}{Marie Babel}, \bibinfo{person}{Julien Bruneau}, {and} \bibinfo{person}{Julien Pettre}.} \bibinfo{year}{2020}\natexlab{}.
\newblock \showarticletitle{Toward Virtual Reality-based Evaluation of Robot Navigation among People}. In \bibinfo{booktitle}{\emph{2020 IEEE Conference on Virtual Reality and 3D User Interfaces (VR)}}. \bibinfo{publisher}{IEEE}, \bibinfo{address}{Atlanta, USA}, \bibinfo{pages}{766--774}.
\newblock
\urldef\tempurl%
\url{https://doi.org/10.1109/VR46266.2020.00100}
\showDOI{\tempurl}


\bibitem[Hakli et~al\mbox{.}(2023)]%
        {hakli2023helping}
\bibfield{author}{\bibinfo{person}{R Hakli} {et~al\mbox{.}}} \bibinfo{year}{2023}\natexlab{}.
\newblock \showarticletitle{Helping-as-Work and Helping-as-Care: Mapping Ambiguities of Helping Commercial Delivery Robots}.
\newblock \bibinfo{journal}{\emph{Social Robots in Social Institutions: Proceedings of Robophilosophy 2022}}  \bibinfo{volume}{366} (\bibinfo{year}{2023}), \bibinfo{pages}{239}.
\newblock
\urldef\tempurl%
\url{https://doi.org/10.3233/FAIA220623}
\showDOI{\tempurl}


\bibitem[Harrison et~al\mbox{.}(2007)]%
        {harrison2007three}
\bibfield{author}{\bibinfo{person}{Steve Harrison}, \bibinfo{person}{Deborah Tatar}, {and} \bibinfo{person}{Phoebe Sengers}.} \bibinfo{year}{2007}\natexlab{}.
\newblock \showarticletitle{The three paradigms of HCI}. In \bibinfo{booktitle}{\emph{Alt. Chi. Session at the SIGCHI Conference on human factors in computing systems San Jose, California, USA}}. \bibinfo{publisher}{ACM}, \bibinfo{address}{California, USA}, \bibinfo{pages}{1--18}.
\newblock


\bibitem[Herdel et~al\mbox{.}(2021)]%
        {Herdel2021Drone}
\bibfield{author}{\bibinfo{person}{Viviane Herdel}, \bibinfo{person}{Anastasia Kuzminykh}, \bibinfo{person}{Andrea Hildebrandt}, {and} \bibinfo{person}{Jessica~R. Cauchard}.} \bibinfo{year}{2021}\natexlab{}.
\newblock \showarticletitle{Drone in Love: Emotional Perception of Facial Expressions on Flying Robots}. In \bibinfo{booktitle}{\emph{Proceedings of the 2021 CHI Conference on Human Factors in Computing Systems}} (<conf-loc>, <city>Yokohama</city>, <country>Japan</country>, </conf-loc>) \emph{(\bibinfo{series}{CHI '21})}. \bibinfo{publisher}{Association for Computing Machinery}, \bibinfo{address}{New York, NY, USA}, Article \bibinfo{articleno}{716}, \bibinfo{numpages}{20}~pages.
\newblock
\showISBNx{9781450380966}
\urldef\tempurl%
\url{https://doi.org/10.1145/3411764.3445495}
\showDOI{\tempurl}


\bibitem[Hoffmann and Prause(2018)]%
        {machines2018Regulatory}
\bibfield{author}{\bibinfo{person}{Thomas Hoffmann} {and} \bibinfo{person}{Gunnar Prause}.} \bibinfo{year}{2018}\natexlab{}.
\newblock \showarticletitle{On the Regulatory Framework for Last-Mile Delivery Robots}.
\newblock \bibinfo{journal}{\emph{Machines}} \bibinfo{volume}{6}, \bibinfo{number}{3} (\bibinfo{year}{2018}), \bibinfo{pages}{1--16}.
\newblock
\showISSN{2075-1702}
\urldef\tempurl%
\url{https://doi.org/10.3390/machines6030033}
\showDOI{\tempurl}


\bibitem[Hoggenmueller et~al\mbox{.}(2020)]%
        {Marius2020Woody}
\bibfield{author}{\bibinfo{person}{Marius Hoggenmueller}, \bibinfo{person}{Luke Hespanhol}, {and} \bibinfo{person}{Martin Tomitsch}.} \bibinfo{year}{2020}\natexlab{}.
\newblock \showarticletitle{Stop and Smell the Chalk Flowers: A Robotic Probe for Investigating Urban Interaction with Physicalised Displays}. In \bibinfo{booktitle}{\emph{Proceedings of the 2020 CHI Conference on Human Factors in Computing Systems}} (Honolulu, HI, USA,) \emph{(\bibinfo{series}{CHI '20})}. \bibinfo{publisher}{Association for Computing Machinery}, \bibinfo{address}{New York, NY, USA}, \bibinfo{pages}{1–14}.
\newblock
\showISBNx{9781450367080}
\urldef\tempurl%
\url{https://doi.org/10.1145/3313831.3376676}
\showDOI{\tempurl}


\bibitem[Hoggenmueller et~al\mbox{.}(2018)]%
        {Marius2018LowRes}
\bibfield{author}{\bibinfo{person}{Marius Hoggenmueller}, \bibinfo{person}{Martin Tomitsch}, {and} \bibinfo{person}{Alexander Wiethoff}.} \bibinfo{year}{2018}\natexlab{}.
\newblock \showarticletitle{Understanding Artefact and Process Challenges for Designing Low-Res Lighting Displays}. In \bibinfo{booktitle}{\emph{Proceedings of the 2018 CHI Conference on Human Factors in Computing Systems}} (<conf-loc>, <city>Montreal QC</city>, <country>Canada</country>, </conf-loc>) \emph{(\bibinfo{series}{CHI '18})}. \bibinfo{publisher}{Association for Computing Machinery}, \bibinfo{address}{New York, NY, USA}, \bibinfo{pages}{1–12}.
\newblock
\showISBNx{9781450356206}
\urldef\tempurl%
\url{https://doi.org/10.1145/3173574.3173833}
\showDOI{\tempurl}


\bibitem[Hoggenm\"{u}ller et~al\mbox{.}(2021a)]%
        {hoggenmueller2021}
\bibfield{author}{\bibinfo{person}{Marius Hoggenm\"{u}ller}, \bibinfo{person}{Wen-Ying Lee}, \bibinfo{person}{Luke Hespanhol}, \bibinfo{person}{Malte Jung}, {and} \bibinfo{person}{Martin Tomitsch}.} \bibinfo{year}{2021}\natexlab{a}.
\newblock \showarticletitle{Eliciting New Perspectives in RtD Studies through Annotated Portfolios: A Case Study of Robotic Artefacts}. In \bibinfo{booktitle}{\emph{Proceedings of the 2021 ACM Designing Interactive Systems Conference}} (Virtual Event, USA) \emph{(\bibinfo{series}{DIS '21})}. \bibinfo{publisher}{Association for Computing Machinery}, \bibinfo{address}{New York, NY, USA}, \bibinfo{pages}{1875–1886}.
\newblock
\showISBNx{9781450384766}
\urldef\tempurl%
\url{https://doi.org/10.1145/3461778.3462134}
\showDOI{\tempurl}


\bibitem[Hoggenm\"{u}ller et~al\mbox{.}(2021b)]%
        {hoggenmueller2021context}
\bibfield{author}{\bibinfo{person}{Marius Hoggenm\"{u}ller}, \bibinfo{person}{Martin Tomitsch}, \bibinfo{person}{Luke Hespanhol}, \bibinfo{person}{Tram Thi~Minh Tran}, \bibinfo{person}{Stewart Worrall}, {and} \bibinfo{person}{Eduardo Nebot}.} \bibinfo{year}{2021}\natexlab{b}.
\newblock \showarticletitle{Context-Based Interface Prototyping: Understanding the Effect of Prototype Representation on User Feedback}. In \bibinfo{booktitle}{\emph{Proceedings of the 2021 CHI Conference on Human Factors in Computing Systems}} (Yokohama, Japan) \emph{(\bibinfo{series}{CHI '21})}. \bibinfo{publisher}{Association for Computing Machinery}, \bibinfo{address}{New York, NY, USA}, Article \bibinfo{articleno}{370}, \bibinfo{numpages}{14}~pages.
\newblock
\showISBNx{9781450380966}
\urldef\tempurl%
\url{https://doi.org/10.1145/3411764.3445159}
\showDOI{\tempurl}


\bibitem[Holm et~al\mbox{.}(2022)]%
        {Holm2022Stuck}
\bibfield{author}{\bibinfo{person}{Daniel~Gahner Holm}, \bibinfo{person}{Rasmus~Peter Junge}, \bibinfo{person}{Mads \O{}stergaard}, \bibinfo{person}{Leon Bodenhagen}, {and} \bibinfo{person}{Oskar Palinko}.} \bibinfo{year}{2022}\natexlab{}.
\newblock \showarticletitle{What Will It Take to Help a Stuck Robot? Exploring Signaling Methods for a Mobile Robot}. In \bibinfo{booktitle}{\emph{Proceedings of the 2022 ACM/IEEE International Conference on Human-Robot Interaction}} \emph{(\bibinfo{series}{HRI '22})}. \bibinfo{publisher}{IEEE Press}, \bibinfo{address}{Sapporo, Hokkaido, Japan}, \bibinfo{pages}{797–801}.
\newblock


\bibitem[Huttenrauch and Eklundh(2003)]%
        {Huttenrauch2003ToHelpOrNot}
\bibfield{author}{\bibinfo{person}{H. Huttenrauch} {and} \bibinfo{person}{K.S. Eklundh}.} \bibinfo{year}{2003}\natexlab{}.
\newblock \showarticletitle{To help or not to help a service robot}. In \bibinfo{booktitle}{\emph{The 12th IEEE International Workshop on Robot and Human Interactive Communication, 2003. Proceedings. ROMAN 2003.}} \bibinfo{publisher}{IEEE}, \bibinfo{address}{California, USA}, \bibinfo{pages}{379--384}.
\newblock
\urldef\tempurl%
\url{https://doi.org/10.1109/ROMAN.2003.1251875}
\showDOI{\tempurl}


\bibitem[Jian et~al\mbox{.}(2000)]%
        {jian2000Trust}
\bibfield{author}{\bibinfo{person}{Jiun-Yin Jian}, \bibinfo{person}{Ann~M Bisantz}, {and} \bibinfo{person}{Colin~G Drury}.} \bibinfo{year}{2000}\natexlab{}.
\newblock \showarticletitle{Foundations for an empirically determined scale of trust in automated systems}.
\newblock \bibinfo{journal}{\emph{International journal of cognitive ergonomics}} \bibinfo{volume}{4}, \bibinfo{number}{1} (\bibinfo{year}{2000}), \bibinfo{pages}{53--71}.
\newblock


\bibitem[Kinzer(2009)]%
        {Tweenbots}
\bibfield{author}{\bibinfo{person}{Kacie Kinzer}.} \bibinfo{year}{Y2009}\natexlab{}.
\newblock \bibinfo{title}{Tweenbots}.
\newblock
\newblock
\urldef\tempurl%
\url{http://www.tweenbots.com/}
\showURL{%
\tempurl}
\newblock
\shownote{Accessed: August,2023}.


\bibitem[Knepper et~al\mbox{.}(2015)]%
        {knepper2015recovering}
\bibfield{author}{\bibinfo{person}{Ross~A Knepper}, \bibinfo{person}{Stefanie Tellex}, \bibinfo{person}{Adrian Li}, \bibinfo{person}{Nicholas Roy}, {and} \bibinfo{person}{Daniela Rus}.} \bibinfo{year}{2015}\natexlab{}.
\newblock \showarticletitle{Recovering from failure by asking for help}.
\newblock \bibinfo{journal}{\emph{Autonomous Robots}}  \bibinfo{volume}{39} (\bibinfo{year}{2015}), \bibinfo{pages}{347--362}.
\newblock
\urldef\tempurl%
\url{https://doi.org/10.1007/s10514-015-9460-1}
\showDOI{\tempurl}


\bibitem[Kuijer and Giaccardi(2018)]%
        {Kuijer2018Coperformance}
\bibfield{author}{\bibinfo{person}{Lenneke Kuijer} {and} \bibinfo{person}{Elisa Giaccardi}.} \bibinfo{year}{2018}\natexlab{}.
\newblock \showarticletitle{Co-Performance: Conceptualizing the Role of Artificial Agency in the Design of Everyday Life}. In \bibinfo{booktitle}{\emph{Proceedings of the 2018 CHI Conference on Human Factors in Computing Systems}} (Montreal QC, Canada) \emph{(\bibinfo{series}{CHI '18})}. \bibinfo{publisher}{Association for Computing Machinery}, \bibinfo{address}{New York, NY, USA}, \bibinfo{pages}{1–13}.
\newblock
\showISBNx{9781450356206}
\urldef\tempurl%
\url{https://doi.org/10.1145/3173574.3173699}
\showDOI{\tempurl}


\bibitem[Kwak et~al\mbox{.}(2018)]%
        {KWAK2018helper}
\bibfield{author}{\bibinfo{person}{Dong-Heon~(Austin) Kwak}, \bibinfo{person}{K.~(Ram) Ramamurthy}, \bibinfo{person}{Derek Nazareth}, {and} \bibinfo{person}{Saerom Lee}.} \bibinfo{year}{2018}\natexlab{}.
\newblock \showarticletitle{The moderating role of helper's high in anchoring process: An empirical investigation in the context of charity website design}.
\newblock \bibinfo{journal}{\emph{Computers in Human Behavior}}  \bibinfo{volume}{84} (\bibinfo{year}{2018}), \bibinfo{pages}{230--244}.
\newblock
\showISSN{0747-5632}
\urldef\tempurl%
\url{https://doi.org/10.1016/j.chb.2018.02.024}
\showDOI{\tempurl}


\bibitem[Kwon et~al\mbox{.}(2018)]%
        {Kwon2018Incapability}
\bibfield{author}{\bibinfo{person}{Minae Kwon}, \bibinfo{person}{Sandy~H. Huang}, {and} \bibinfo{person}{Anca~D. Dragan}.} \bibinfo{year}{2018}\natexlab{}.
\newblock \showarticletitle{Expressing Robot Incapability}. In \bibinfo{booktitle}{\emph{Proceedings of the 2018 ACM/IEEE International Conference on Human-Robot Interaction}} (Chicago, IL, USA) \emph{(\bibinfo{series}{HRI '18})}. \bibinfo{publisher}{Association for Computing Machinery}, \bibinfo{address}{New York, NY, USA}, \bibinfo{pages}{87–95}.
\newblock
\showISBNx{9781450349536}
\urldef\tempurl%
\url{https://doi.org/10.1145/3171221.3171276}
\showDOI{\tempurl}


\bibitem[Lee et~al\mbox{.}(2020)]%
        {Lee2019Bubble}
\bibfield{author}{\bibinfo{person}{Wen-Ying Lee}, \bibinfo{person}{Yoyo Tsung-Yu Hou}, \bibinfo{person}{Cristina~Zaga Human}, {and} \bibinfo{person}{Malte Jung}.} \bibinfo{year}{2020}\natexlab{}.
\newblock \showarticletitle{Design for serendipitous interaction: BubbleBot - bringing people together with bubbles}. In \bibinfo{booktitle}{\emph{Proceedings of the 14th ACM/IEEE International Conference on Human-Robot Interaction}} \emph{(\bibinfo{series}{HRI '19})}. \bibinfo{publisher}{IEEE Press}, \bibinfo{address}{Daegu, Republic of Korea}, \bibinfo{pages}{759–760}.
\newblock
\showISBNx{9781538685556}


\bibitem[Lee and Jung(2020)]%
        {Lee2020Ludic}
\bibfield{author}{\bibinfo{person}{Wen-Ying Lee} {and} \bibinfo{person}{Malte Jung}.} \bibinfo{year}{2020}\natexlab{}.
\newblock \showarticletitle{Ludic-HRI: Designing Playful Experiences with Robots}. In \bibinfo{booktitle}{\emph{Companion of the 2020 ACM/IEEE International Conference on Human-Robot Interaction}} (Cambridge, United Kingdom) \emph{(\bibinfo{series}{HRI '20})}. \bibinfo{publisher}{Association for Computing Machinery}, \bibinfo{address}{New York, NY, USA}, \bibinfo{pages}{582–584}.
\newblock
\showISBNx{9781450370578}
\urldef\tempurl%
\url{https://doi.org/10.1145/3371382.3377429}
\showDOI{\tempurl}


\bibitem[Legler et~al\mbox{.}(2023)]%
        {robotics12060168VR}
\bibfield{author}{\bibinfo{person}{Franziska Legler}, \bibinfo{person}{Jonas Trezl}, \bibinfo{person}{Dorothea Langer}, \bibinfo{person}{Max Bernhagen}, \bibinfo{person}{Andre Dettmann}, {and} \bibinfo{person}{Angelika~C. Bullinger}.} \bibinfo{year}{2023}\natexlab{}.
\newblock \showarticletitle{Emotional Experience in Human–Robot Collaboration: Suitability of Virtual Reality Scenarios to Study Interactions beyond Safety Restrictions}.
\newblock \bibinfo{journal}{\emph{Robotics}} \bibinfo{volume}{12}, \bibinfo{number}{6} (\bibinfo{year}{2023}), \bibinfo{pages}{1--23}.
\newblock
\showISSN{2218-6581}
\urldef\tempurl%
\url{https://doi.org/10.3390/robotics12060168}
\showDOI{\tempurl}


\bibitem[Liang et~al\mbox{.}(2023)]%
        {Liang2023Direction}
\bibfield{author}{\bibinfo{person}{Claire Liang}, \bibinfo{person}{Andy~Elliot Ricci}, \bibinfo{person}{Hadas Kress-Gazit}, {and} \bibinfo{person}{Malte~F. Jung}.} \bibinfo{year}{2023}\natexlab{}.
\newblock \showarticletitle{Lessons From a Robot Asking for Directions In-the-Wild}. In \bibinfo{booktitle}{\emph{Companion of the 2023 ACM/IEEE International Conference on Human-Robot Interaction}} (Stockholm, Sweden) \emph{(\bibinfo{series}{HRI '23})}. \bibinfo{publisher}{Association for Computing Machinery}, \bibinfo{address}{New York, NY, USA}, \bibinfo{pages}{617–620}.
\newblock
\showISBNx{9781450399708}
\urldef\tempurl%
\url{https://doi.org/10.1145/3568294.3580159}
\showDOI{\tempurl}


\bibitem[Lorish and Maisiak(1986)]%
        {lorish1986face}
\bibfield{author}{\bibinfo{person}{Christopher~D Lorish} {and} \bibinfo{person}{Richard Maisiak}.} \bibinfo{year}{1986}\natexlab{}.
\newblock \showarticletitle{The face scale: a brief, nonverbal method for assessing patient mood}.
\newblock \bibinfo{journal}{\emph{Arthritis \& Rheumatism: Official Journal of the American College of Rheumatology}} \bibinfo{volume}{29}, \bibinfo{number}{7} (\bibinfo{year}{1986}), \bibinfo{pages}{906--909}.
\newblock
\urldef\tempurl%
\url{https://doi.org/10.1002/art.1780290714}
\showDOI{\tempurl}


\bibitem[Lupetti(2016)]%
        {Lupetti2016}
\bibfield{author}{\bibinfo{person}{Maria~Luce Lupetti}.} \bibinfo{year}{2016}\natexlab{}.
\newblock \showarticletitle{Designing Playful HRI: Acceptability of Robots in Everyday Life through Play}. In \bibinfo{booktitle}{\emph{The Eleventh ACM/IEEE International Conference on Human Robot Interaction}} \emph{(\bibinfo{series}{HRI '16})}. \bibinfo{publisher}{IEEE Press}, \bibinfo{address}{Christchurch, New Zealand}, \bibinfo{pages}{631–632}.
\newblock
\showISBNx{9781467383707}


\bibitem[Lupetti(2020)]%
        {Lupetti_2020}
\bibfield{author}{\bibinfo{person}{Maria~Luce Lupetti}.} \bibinfo{year}{2020}\natexlab{}.
\newblock \showarticletitle{Shybo – Design of a research artefact for human-robot interaction studies}.
\newblock \bibinfo{journal}{\emph{Journal of Science and Technology of the Arts}} \bibinfo{volume}{9}, \bibinfo{number}{1} (\bibinfo{date}{Mar.} \bibinfo{year}{2020}), \bibinfo{pages}{57--69}.
\newblock
\urldef\tempurl%
\url{https://doi.org/10.7559/citarj.v9i1.303}
\showDOI{\tempurl}


\bibitem[Lupetti et~al\mbox{.}(2019)]%
        {lupetti2019citizenship}
\bibfield{author}{\bibinfo{person}{Maria~Luce Lupetti}, \bibinfo{person}{Roy Bendor}, {and} \bibinfo{person}{Elisa Giaccardi}.} \bibinfo{year}{2019}\natexlab{}.
\newblock \showarticletitle{Robot citizenship: A design perspective}.
\newblock \bibinfo{journal}{\emph{Design and Semantics of Form and Movement}}  \bibinfo{volume}{87} (\bibinfo{year}{2019}), \bibinfo{pages}{81--89}.
\newblock


\bibitem[Lupetti et~al\mbox{.}(2018)]%
        {mti2040069}
\bibfield{author}{\bibinfo{person}{Maria~Luce Lupetti}, \bibinfo{person}{Giovanni Piumatti}, \bibinfo{person}{Claudio Germak}, {and} \bibinfo{person}{Fabrizio Lamberti}.} \bibinfo{year}{2018}\natexlab{}.
\newblock \showarticletitle{Design and Evaluation of a Mixed-Reality Playground for Child-Robot Games}.
\newblock \bibinfo{journal}{\emph{Multimodal Technologies and Interaction}} \bibinfo{volume}{2}, \bibinfo{number}{4} (\bibinfo{year}{2018}), \bibinfo{pages}{1--16}.
\newblock
\showISSN{2414-4088}
\urldef\tempurl%
\url{https://doi.org/10.3390/mti2040069}
\showDOI{\tempurl}


\bibitem[Lupetti and Van~Mechelen(2022)]%
        {Lupetti2022}
\bibfield{author}{\bibinfo{person}{Maria~Luce Lupetti} {and} \bibinfo{person}{Maarten Van~Mechelen}.} \bibinfo{year}{2022}\natexlab{}.
\newblock \showarticletitle{Promoting Children's Critical Thinking Towards Robotics through Robot Deception}. In \bibinfo{booktitle}{\emph{Proceedings of the 2022 ACM/IEEE International Conference on Human-Robot Interaction}} \emph{(\bibinfo{series}{HRI '22})}. \bibinfo{publisher}{IEEE Press}, \bibinfo{address}{Sapporo, Hokkaido, Japan}, \bibinfo{pages}{588–597}.
\newblock


\bibitem[Mail(2023)]%
        {dailymail_2023}
\bibfield{author}{\bibinfo{person}{Daily Mail}.} \bibinfo{year}{2023}\natexlab{}.
\newblock \bibinfo{title}{Pictured: Delivery robots queue patiently to use pedestrian crossing in Cambridge}.
\newblock
\newblock
\urldef\tempurl%
\url{https://www.dailymail.co.uk/sciencetech/article-11497909/Pictured-Delivery-robots-queue-patiently-use-pedestrian-crossing-Cambridge.html}
\showURL{%
\tempurl}
\newblock
\shownote{Accessed on January 28, 2024}.


\bibitem[Malle and Ullman(2023)]%
        {malle2023trust}
\bibfield{author}{\bibinfo{person}{Bertram~F Malle} {and} \bibinfo{person}{Daniel Ullman}.} \bibinfo{year}{2023}\natexlab{}.
\newblock \showarticletitle{Measuring Human-Robot Trust with the MDMT (Multi-Dimensional Measure of Trust)}.
\newblock \bibinfo{journal}{\emph{arXiv preprint arXiv:2311.14887}}  \bibinfo{volume}{null} (\bibinfo{year}{2023}), \bibinfo{pages}{1--3}.
\newblock


\bibitem[Marenko and Van~Allen(2016)]%
        {marenko2016animistic}
\bibfield{author}{\bibinfo{person}{Betti Marenko} {and} \bibinfo{person}{Philip Van~Allen}.} \bibinfo{year}{2016}\natexlab{}.
\newblock \showarticletitle{Animistic design: how to reimagine digital interaction between the human and the nonhuman}.
\newblock \bibinfo{journal}{\emph{Digital Creativity}} \bibinfo{volume}{27}, \bibinfo{number}{1} (\bibinfo{year}{2016}), \bibinfo{pages}{52--70}.
\newblock
\urldef\tempurl%
\url{https://doi.org/10.1080/14626268.2016.1145127}
\showDOI{\tempurl}


\bibitem[McAllister(1995)]%
        {McAllister1995AffectAC}
\bibfield{author}{\bibinfo{person}{Daniel~J. McAllister}.} \bibinfo{year}{1995}\natexlab{}.
\newblock \showarticletitle{Affect- and Cognition-Based Trust as Foundations for Interpersonal Cooperation in Organizations}.
\newblock \bibinfo{journal}{\emph{Academy of Management Journal}}  \bibinfo{volume}{38} (\bibinfo{year}{1995}), \bibinfo{pages}{24--59}.
\newblock
\urldef\tempurl%
\url{https://api.semanticscholar.org/CorpusID:32041515}
\showURL{%
\tempurl}


\bibitem[Milde et~al\mbox{.}(2023)]%
        {Milde2023MultiModalVR}
\bibfield{author}{\bibinfo{person}{Sven Milde}, \bibinfo{person}{Tabea Runzheimer}, \bibinfo{person}{Stefan Friesen}, \bibinfo{person}{Johannes-Hubert Peiffer}, \bibinfo{person}{Johannes-Jeremias H{\"o}fler}, \bibinfo{person}{Kerstin Geis}, \bibinfo{person}{Jan-Torsten Milde}, {and} \bibinfo{person}{Rainer Blum}.} \bibinfo{year}{2023}\natexlab{}.
\newblock \showarticletitle{Studying Multi-modal Human Robot Interaction Using a Mobile VR Simulation}. In \bibinfo{booktitle}{\emph{Human-Computer Interaction}}, \bibfield{editor}{\bibinfo{person}{Masaaki Kurosu} {and} \bibinfo{person}{Ayako Hashizume}} (Eds.). \bibinfo{publisher}{Springer Nature Switzerland}, \bibinfo{address}{Cham}, \bibinfo{pages}{140--155}.
\newblock
\showISBNx{978-3-031-35602-5}


\bibitem[Nanavati et~al\mbox{.}(2021)]%
        {nanavati2021modelingHelpfulness}
\bibfield{author}{\bibinfo{person}{Amal Nanavati}, \bibinfo{person}{Christoforos~I Mavrogiannis}, \bibinfo{person}{Kevin Weatherwax}, \bibinfo{person}{Leila Takayama}, \bibinfo{person}{Maya Cakmak}, {and} \bibinfo{person}{Siddhartha~S Srinivasa}.} \bibinfo{year}{2021}\natexlab{}.
\newblock \showarticletitle{Modeling Human Helpfulness with Individual and Contextual Factors for Robot Planning.}. In \bibinfo{booktitle}{\emph{Robotics: Science and Systems}}. \bibinfo{publisher}{Robotics: Science and Systems Foundation}, \bibinfo{address}{virtual}, \bibinfo{pages}{1--11}.
\newblock


\bibitem[Nguyen et~al\mbox{.}(2023)]%
        {Nguyen2023}
\bibfield{author}{\bibinfo{person}{Binh Vinh~Duc Nguyen}, \bibinfo{person}{Jihae Han}, \bibinfo{person}{Maarten Houben}, \bibinfo{person}{Yssmin Bayoumi}, {and} \bibinfo{person}{Andrew Vande~Moere}.} \bibinfo{year}{2023}\natexlab{}.
\newblock \showarticletitle{Engaging Passers-by with Rhythm: Applying Feedforward Learning to a Xylophonic Media Architecture Facade}. In \bibinfo{booktitle}{\emph{Proceedings of the 2023 CHI Conference on Human Factors in Computing Systems}} (<conf-loc>, <city>Hamburg</city>, <country>Germany</country>, </conf-loc>) \emph{(\bibinfo{series}{CHI '23})}. \bibinfo{publisher}{Association for Computing Machinery}, \bibinfo{address}{New York, NY, USA}, Article \bibinfo{articleno}{182}, \bibinfo{numpages}{21}~pages.
\newblock
\showISBNx{9781450394215}
\urldef\tempurl%
\url{https://doi.org/10.1145/3544548.3580761}
\showDOI{\tempurl}


\bibitem[Palmipuu(2021)]%
        {postimees2021robotsnow}
\bibfield{author}{\bibinfo{person}{Triin Palmipuu}.} \bibinfo{year}{2021}\natexlab{}.
\newblock \bibinfo{title}{Pelgulinlane Taivo aitab iga päev pakiroboteid: nad paluvad nii härdalt abi}.
\newblock
\newblock
\urldef\tempurl%
\url{https://naine.postimees.ee/7406578/pelgulinlane-taivo-aitab-iga-paevpakiroboteid-nad-paluvad-nii-hardalt-abi}
\showURL{%
\tempurl}
\newblock
\shownote{Accessed:2023}.


\bibitem[Pelikan et~al\mbox{.}(2024)]%
        {pelikan2024encountering}
\bibfield{author}{\bibinfo{person}{Hannah R.~M. Pelikan}, \bibinfo{person}{Stuart Reeves}, {and} \bibinfo{person}{Marina~N. Cantarutti}.} \bibinfo{year}{2024}\natexlab{}.
\newblock \showarticletitle{Encountering Autonomous Robots on Public Streets}. In \bibinfo{booktitle}{\emph{Proceedings of the 2024 ACM/IEEE International Conference on Human-Robot Interaction}} (<conf-loc>, <city>Boulder</city>, <state>CO</state>, <country>USA</country>, </conf-loc>) \emph{(\bibinfo{series}{HRI '24})}. \bibinfo{publisher}{Association for Computing Machinery}, \bibinfo{address}{New York, NY, USA}, \bibinfo{pages}{561–571}.
\newblock
\showISBNx{9798400703225}
\urldef\tempurl%
\url{https://doi.org/10.1145/3610977.3634936}
\showDOI{\tempurl}


\bibitem[Riedmann et~al\mbox{.}(2022)]%
        {Riedmann2022}
\bibfield{author}{\bibinfo{person}{Anna Riedmann}, \bibinfo{person}{Philipp Schaper}, {and} \bibinfo{person}{Birgit Lugrin}.} \bibinfo{year}{2022}\natexlab{}.
\newblock \showarticletitle{Integration of a social robot and gamification in adult learning and effects on motivation, engagement and performance}.
\newblock \bibinfo{journal}{\emph{AI {\&} SOCIETY}} \bibinfo{volume}{1}, \bibinfo{number}{39} (\bibinfo{date}{25 Jun} \bibinfo{year}{2022}), \bibinfo{pages}{369--388}.
\newblock
\showISSN{1435-5655}
\urldef\tempurl%
\url{https://doi.org/10.1007/s00146-022-01514-y}
\showDOI{\tempurl}


\bibitem[Robson et~al\mbox{.}(2015)]%
        {ROBSON2015411GamePrinciples}
\bibfield{author}{\bibinfo{person}{Karen Robson}, \bibinfo{person}{Kirk Plangger}, \bibinfo{person}{Jan~H. Kietzmann}, \bibinfo{person}{Ian McCarthy}, {and} \bibinfo{person}{Leyland Pitt}.} \bibinfo{year}{2015}\natexlab{}.
\newblock \showarticletitle{Is it all a game? Understanding the principles of gamification}.
\newblock \bibinfo{journal}{\emph{Business Horizons}} \bibinfo{volume}{58}, \bibinfo{number}{4} (\bibinfo{year}{2015}), \bibinfo{pages}{411--420}.
\newblock
\showISSN{0007-6813}
\urldef\tempurl%
\url{https://doi.org/10.1016/j.bushor.2015.03.006}
\showDOI{\tempurl}


\bibitem[Rosenthal et~al\mbox{.}(2012)]%
        {rosenthal2012someone}
\bibfield{author}{\bibinfo{person}{Stephanie Rosenthal}, \bibinfo{person}{Manuela Veloso}, {and} \bibinfo{person}{Anind~K Dey}.} \bibinfo{year}{2012}\natexlab{}.
\newblock \showarticletitle{Is someone in this office available to help me? Proactively seeking help from spatially-situated humans}.
\newblock \bibinfo{journal}{\emph{Journal of Intelligent \& Robotic Systems}}  \bibinfo{volume}{66} (\bibinfo{year}{2012}), \bibinfo{pages}{205--221}.
\newblock
\urldef\tempurl%
\url{https://doi.org/10.1007/s10846-011-9610-4}
\showDOI{\tempurl}


\bibitem[Rosenthal-von~der P\"{u}tten et~al\mbox{.}(2020)]%
        {Putten2020Forgotten}
\bibfield{author}{\bibinfo{person}{Astrid Rosenthal-von~der P\"{u}tten}, \bibinfo{person}{David Sirkin}, \bibinfo{person}{Anna Abrams}, {and} \bibinfo{person}{Laura Platte}.} \bibinfo{year}{2020}\natexlab{}.
\newblock \showarticletitle{The Forgotten in HRI: Incidental Encounters with Robots in Public Spaces}. In \bibinfo{booktitle}{\emph{Companion of the 2020 ACM/IEEE International Conference on Human-Robot Interaction}} (Cambridge, United Kingdom) \emph{(\bibinfo{series}{HRI '20})}. \bibinfo{publisher}{Association for Computing Machinery}, \bibinfo{address}{New York, NY, USA}, \bibinfo{pages}{656–657}.
\newblock
\showISBNx{9781450370578}
\urldef\tempurl%
\url{https://doi.org/10.1145/3371382.3374852}
\showDOI{\tempurl}


\bibitem[Salvini(2018)]%
        {SALVINI2018UrbanRobot}
\bibfield{author}{\bibinfo{person}{Pericle Salvini}.} \bibinfo{year}{2018}\natexlab{}.
\newblock \showarticletitle{Urban robotics: Towards responsible innovations for our cities}.
\newblock \bibinfo{journal}{\emph{Robotics and Autonomous Systems}}  \bibinfo{volume}{100} (\bibinfo{year}{2018}), \bibinfo{pages}{278--286}.
\newblock
\showISSN{0921-8890}
\urldef\tempurl%
\url{https://doi.org/10.1016/j.robot.2017.03.007}
\showDOI{\tempurl}


\bibitem[Salvini et~al\mbox{.}(2010)]%
        {Bully2010Salvini}
\bibfield{author}{\bibinfo{person}{P. Salvini}, \bibinfo{person}{G. Ciaravella}, \bibinfo{person}{W. Yu}, \bibinfo{person}{G. Ferri}, \bibinfo{person}{A. Manzi}, \bibinfo{person}{B. Mazzolai}, \bibinfo{person}{C. Laschi}, \bibinfo{person}{S.R. Oh}, {and} \bibinfo{person}{P. Dario}.} \bibinfo{year}{2010}\natexlab{}.
\newblock \showarticletitle{How safe are service robots in urban environments? Bullying a robot}. In \bibinfo{booktitle}{\emph{19th International Symposium in Robot and Human Interactive Communication}}. \bibinfo{publisher}{IEEE}, \bibinfo{address}{Viareggio, Italy}, \bibinfo{pages}{1--7}.
\newblock
\urldef\tempurl%
\url{https://doi.org/10.1109/ROMAN.2010.5654677}
\showDOI{\tempurl}


\bibitem[Saupp\'{e} and Mutlu(2015)]%
        {Saupp2015Industry}
\bibfield{author}{\bibinfo{person}{Allison Saupp\'{e}} {and} \bibinfo{person}{Bilge Mutlu}.} \bibinfo{year}{2015}\natexlab{}.
\newblock \showarticletitle{The Social Impact of a Robot Co-Worker in Industrial Settings}. In \bibinfo{booktitle}{\emph{Proceedings of the 33rd Annual ACM Conference on Human Factors in Computing Systems}} (Seoul, Republic of Korea) \emph{(\bibinfo{series}{CHI '15})}. \bibinfo{publisher}{Association for Computing Machinery}, \bibinfo{address}{New York, NY, USA}, \bibinfo{pages}{3613–3622}.
\newblock
\showISBNx{9781450331456}
\urldef\tempurl%
\url{https://doi.org/10.1145/2702123.2702181}
\showDOI{\tempurl}


\bibitem[Schneiders et~al\mbox{.}(2021)]%
        {Schneiders2021Domestic}
\bibfield{author}{\bibinfo{person}{Eike Schneiders}, \bibinfo{person}{Anne~Marie Kanstrup}, \bibinfo{person}{Jesper Kjeldskov}, {and} \bibinfo{person}{Mikael~B. Skov}.} \bibinfo{year}{2021}\natexlab{}.
\newblock \showarticletitle{Domestic Robots and the Dream of Automation: Understanding Human Interaction and Intervention}. In \bibinfo{booktitle}{\emph{Proceedings of the 2021 CHI Conference on Human Factors in Computing Systems}} (<conf-loc>, <city>Yokohama</city>, <country>Japan</country>, </conf-loc>) \emph{(\bibinfo{series}{CHI '21})}. \bibinfo{publisher}{Association for Computing Machinery}, \bibinfo{address}{New York, NY, USA}, Article \bibinfo{articleno}{241}, \bibinfo{numpages}{13}~pages.
\newblock
\showISBNx{9781450380966}
\urldef\tempurl%
\url{https://doi.org/10.1145/3411764.3445629}
\showDOI{\tempurl}


\bibitem[Schrepp et~al\mbox{.}(2017)]%
        {Schrepp2017UEQS}
\bibfield{author}{\bibinfo{person}{Martin Schrepp}, \bibinfo{person}{Andreas Hinderks}, {and} \bibinfo{person}{J\"{o}rg Thomaschewski}.} \bibinfo{year}{2017}\natexlab{}.
\newblock \showarticletitle{Design and evaluation of a short version of the User Experience Questionnaire (UEQ-S)}.
\newblock \bibinfo{journal}{\emph{International Journal of Interactive Multimedia and Artificial Intelligence}} \bibinfo{volume}{4}, \bibinfo{number}{6} (\bibinfo{year}{2017}), \bibinfo{pages}{103}.
\newblock
\urldef\tempurl%
\url{https://doi.org/10.9781/ijimai.2017.09.001}
\showDOI{\tempurl}


\bibitem[Sirkin et~al\mbox{.}(2015)]%
        {Ottoman2015Sirkin}
\bibfield{author}{\bibinfo{person}{David Sirkin}, \bibinfo{person}{Brian Mok}, \bibinfo{person}{Stephen Yang}, {and} \bibinfo{person}{Wendy Ju}.} \bibinfo{year}{2015}\natexlab{}.
\newblock \showarticletitle{Mechanical Ottoman: How Robotic Furniture Offers and Withdraws Support}. In \bibinfo{booktitle}{\emph{Proceedings of the Tenth Annual ACM/IEEE International Conference on Human-Robot Interaction}} (Portland, Oregon, USA) \emph{(\bibinfo{series}{HRI '15})}. \bibinfo{publisher}{Association for Computing Machinery}, \bibinfo{address}{New York, NY, USA}, \bibinfo{pages}{11–18}.
\newblock
\showISBNx{9781450328838}
\urldef\tempurl%
\url{https://doi.org/10.1145/2696454.2696461}
\showDOI{\tempurl}


\bibitem[Smith and Zeller(2017)]%
        {Smith2017Hitchbot}
\bibfield{author}{\bibinfo{person}{David~Harris Smith} {and} \bibinfo{person}{Frauke Zeller}.} \bibinfo{year}{2017}\natexlab{}.
\newblock \showarticletitle{{The Death and Lives of hitchBOT: The Design and Implementation of a Hitchhiking Robot}}.
\newblock \bibinfo{journal}{\emph{Leonardo}} \bibinfo{volume}{50}, \bibinfo{number}{1} (\bibinfo{date}{02} \bibinfo{year}{2017}), \bibinfo{pages}{77--78}.
\newblock
\showISSN{0024-094X}
\urldef\tempurl%
\url{https://doi.org/10.1162/LEON_a_01354}
\showDOI{\tempurl}


\bibitem[Srinivasan and Takayama(2016)]%
        {Srinivasan2016HelpMePlease}
\bibfield{author}{\bibinfo{person}{Vasant Srinivasan} {and} \bibinfo{person}{Leila Takayama}.} \bibinfo{year}{2016}\natexlab{}.
\newblock \showarticletitle{Help Me Please: Robot Politeness Strategies for Soliciting Help From Humans}. In \bibinfo{booktitle}{\emph{Proceedings of the 2016 CHI Conference on Human Factors in Computing Systems}} (San Jose, California, USA) \emph{(\bibinfo{series}{CHI '16})}. \bibinfo{publisher}{Association for Computing Machinery}, \bibinfo{address}{New York, NY, USA}, \bibinfo{pages}{4945–4955}.
\newblock
\showISBNx{9781450333627}
\urldef\tempurl%
\url{https://doi.org/10.1145/2858036.2858217}
\showDOI{\tempurl}


\bibitem[Tan et~al\mbox{.}(2018)]%
        {Abuse2018Tan}
\bibfield{author}{\bibinfo{person}{Xiang~Zhi Tan}, \bibinfo{person}{Marynel V\'{a}zquez}, \bibinfo{person}{Elizabeth~J. Carter}, \bibinfo{person}{Cecilia~G. Morales}, {and} \bibinfo{person}{Aaron Steinfeld}.} \bibinfo{year}{2018}\natexlab{}.
\newblock \showarticletitle{Inducing Bystander Interventions During Robot Abuse with Social Mechanisms}. In \bibinfo{booktitle}{\emph{Proceedings of the 2018 ACM/IEEE International Conference on Human-Robot Interaction}} \emph{(\bibinfo{series}{HRI '18})}. \bibinfo{publisher}{Association for Computing Machinery}, \bibinfo{address}{New York, NY, USA}, \bibinfo{pages}{169–177}.
\newblock
\showISBNx{9781450349536}
\urldef\tempurl%
\url{https://doi.org/10.1145/3171221.3171247}
\showDOI{\tempurl}


\bibitem[Tomitsch(2014)]%
        {tomitsch2014PublicDisplay}
\bibfield{author}{\bibinfo{person}{Martin Tomitsch}.} \bibinfo{year}{2014}\natexlab{}.
\newblock \showarticletitle{Towards the real-time city: An investigation of public displays for behaviour change and sustainable living}. In \bibinfo{booktitle}{\emph{7th making cities livable conference}}. \bibinfo{publisher}{Association for Sustainability in Business}, \bibinfo{address}{NSW, Australia}, \bibinfo{pages}{1--19}.
\newblock


\bibitem[Tondello(2016)]%
        {Tondello2016}
\bibfield{author}{\bibinfo{person}{Gustavo~Fortes Tondello}.} \bibinfo{year}{2016}\natexlab{}.
\newblock \showarticletitle{An introduction to gamification in human-computer interaction}.
\newblock \bibinfo{journal}{\emph{XRDS}} \bibinfo{volume}{23}, \bibinfo{number}{1} (\bibinfo{date}{sep} \bibinfo{year}{2016}), \bibinfo{pages}{15–17}.
\newblock
\showISSN{1528-4972}
\urldef\tempurl%
\url{https://doi.org/10.1145/2983457}
\showDOI{\tempurl}


\bibitem[Urakami(2023)]%
        {urakami2023emotional}
\bibfield{author}{\bibinfo{person}{Jacqueline Urakami}.} \bibinfo{year}{2023}\natexlab{}.
\newblock \showarticletitle{Do Emotional Robots Get More Help? How a Robots Emotions Affect Collaborators Willingness to Help}.
\newblock \bibinfo{journal}{\emph{International Journal of Social Robotics}} \bibinfo{volume}{15}, \bibinfo{number}{9} (\bibinfo{year}{2023}), \bibinfo{pages}{1457--1471}.
\newblock
\urldef\tempurl%
\url{https://doi.org/10.1007/s12369-023-01058-1}
\showDOI{\tempurl}


\bibitem[van~der Grinten et~al\mbox{.}(2020)]%
        {van2020designing}
\bibfield{author}{\bibinfo{person}{Tim van~der Grinten}, \bibinfo{person}{Steffen M{\"u}ller}, \bibinfo{person}{Martin Westhoven}, \bibinfo{person}{Sascha Wischniewski}, \bibinfo{person}{Andrea Scheidig}, {and} \bibinfo{person}{Horst-Michael Gross}.} \bibinfo{year}{2020}\natexlab{}.
\newblock \showarticletitle{Designing an expressive head for a help requesting socially assistive robot}. In \bibinfo{booktitle}{\emph{Human-Friendly Robotics 2019: 12th International Workshop}}. \bibinfo{publisher}{Springer}, \bibinfo{address}{Reggio Emilia, Italy}, \bibinfo{pages}{88--102}.
\newblock
\urldef\tempurl%
\url{https://doi.org/10.1007/978-3-030-42026-0_7}
\showDOI{\tempurl}


\bibitem[{Van Der Laan} et~al\mbox{.}(1997)]%
        {VANDERLAAN19971Acceptance}
\bibfield{author}{\bibinfo{person}{Jinke~D. {Van Der Laan}}, \bibinfo{person}{Adriaan Heino}, {and} \bibinfo{person}{Dick {De Waard}}.} \bibinfo{year}{1997}\natexlab{}.
\newblock \showarticletitle{A simple procedure for the assessment of acceptance of advanced transport telematics}.
\newblock \bibinfo{journal}{\emph{Transportation Research Part C: Emerging Technologies}} \bibinfo{volume}{5}, \bibinfo{number}{1} (\bibinfo{year}{1997}), \bibinfo{pages}{1--10}.
\newblock
\showISSN{0968-090X}
\urldef\tempurl%
\url{https://doi.org/10.1016/S0968-090X(96)00025-3}
\showDOI{\tempurl}


\bibitem[Veloso(2018)]%
        {Veloso2018Opportunity}
\bibfield{author}{\bibinfo{person}{Manuela~M. Veloso}.} \bibinfo{year}{2018}\natexlab{}.
\newblock \showarticletitle{The Increasingly Fascinating Opportunity for Human-Robot-AI Interaction: The CoBot Mobile Service Robots}.
\newblock \bibinfo{journal}{\emph{J. Hum.-Robot Interact.}} \bibinfo{volume}{7}, \bibinfo{number}{1}, Article \bibinfo{articleno}{5} (\bibinfo{date}{may} \bibinfo{year}{2018}), \bibinfo{numpages}{2}~pages.
\newblock
\urldef\tempurl%
\url{https://doi.org/10.1145/3209541}
\showDOI{\tempurl}


\bibitem[Venås et~al\mbox{.}(2024)]%
        {Venas2024}
\bibfield{author}{\bibinfo{person}{Gizem~Ateş Venås}, \bibinfo{person}{Martin~Fodstad Stølen}, {and} \bibinfo{person}{Erik Kyrkjebø}.} \bibinfo{year}{2024}\natexlab{}.
\newblock \showarticletitle{Exploring human-robot cooperation with gamified user training: a user study on cooperative lifting}.
\newblock \bibinfo{journal}{\emph{Frontiers in Robotics and AI}}  \bibinfo{volume}{10} (\bibinfo{year}{2024}), \bibinfo{pages}{1--20}.
\newblock
\showISSN{2296-9144}
\urldef\tempurl%
\url{https://doi.org/10.3389/frobt.2023.1290104}
\showDOI{\tempurl}


\bibitem[Villani et~al\mbox{.}(2018)]%
        {Villani2018VR}
\bibfield{author}{\bibinfo{person}{Valeria Villani}, \bibinfo{person}{Beatrice Capelli}, {and} \bibinfo{person}{Lorenzo Sabattini}.} \bibinfo{year}{2018}\natexlab{}.
\newblock \showarticletitle{Use of Virtual Reality for the Evaluation of Human-Robot Interaction Systems in Complex Scenarios}. In \bibinfo{booktitle}{\emph{2018 27th IEEE International Symposium on Robot and Human Interactive Communication (RO-MAN)}}. \bibinfo{publisher}{IEEE}, \bibinfo{address}{Nanjing, China}, \bibinfo{pages}{422--427}.
\newblock
\urldef\tempurl%
\url{https://doi.org/10.1109/ROMAN.2018.8525738}
\showDOI{\tempurl}


\bibitem[Weinberg et~al\mbox{.}(2023)]%
        {Weinberg2023Observe}
\bibfield{author}{\bibinfo{person}{David Weinberg}, \bibinfo{person}{Healy Dwyer}, \bibinfo{person}{Sarah~E. Fox}, {and} \bibinfo{person}{Nikolas Martelaro}.} \bibinfo{year}{2023}\natexlab{}.
\newblock \showarticletitle{Sharing the Sidewalk: Observing Delivery Robot Interactions with Pedestrians during a Pilot in Pittsburgh, PA}.
\newblock \bibinfo{journal}{\emph{Multimodal Technologies and Interaction}} \bibinfo{volume}{7}, \bibinfo{number}{5} (\bibinfo{year}{2023}), \bibinfo{pages}{1--27}.
\newblock
\showISSN{2414-4088}
\urldef\tempurl%
\url{https://doi.org/10.3390/mti7050053}
\showDOI{\tempurl}


\bibitem[Weiss et~al\mbox{.}(2010)]%
        {Weiss2010ACE}
\bibfield{author}{\bibinfo{person}{Astrid Weiss}, \bibinfo{person}{Judith Igelsb\"{o}ck}, \bibinfo{person}{Manfred Tscheligi}, \bibinfo{person}{Andrea Bauer}, \bibinfo{person}{Kolja K\"{u}hnlenz}, \bibinfo{person}{Dirk Wollherr}, {and} \bibinfo{person}{Martin Buss}.} \bibinfo{year}{2010}\natexlab{}.
\newblock \showarticletitle{Robots asking for directions: the willingness of passers-by to support robots}. In \bibinfo{booktitle}{\emph{Proceedings of the 5th ACM/IEEE International Conference on Human-Robot Interaction}} \emph{(\bibinfo{series}{HRI '10})}. \bibinfo{publisher}{IEEE Press}, \bibinfo{address}{Osaka, Japan}, \bibinfo{pages}{23–30}.
\newblock
\showISBNx{9781424448937}


\bibitem[Yamaji et~al\mbox{.}(2010)]%
        {TrashBin2010Children}
\bibfield{author}{\bibinfo{person}{Yuto Yamaji}, \bibinfo{person}{Taisuke Miyake}, \bibinfo{person}{Yuta Yoshiike}, \bibinfo{person}{P.~Ravindra~S. De~Silva}, {and} \bibinfo{person}{Michio Okada}.} \bibinfo{year}{2010}\natexlab{}.
\newblock \showarticletitle{STB: human-dependent sociable trash box}. In \bibinfo{booktitle}{\emph{Proceedings of the 5th ACM/IEEE International Conference on Human-Robot Interaction}} \emph{(\bibinfo{series}{HRI '10})}. \bibinfo{publisher}{IEEE Press}, \bibinfo{address}{Osaka, Japan}, \bibinfo{pages}{197–198}.
\newblock
\showISBNx{9781424448937}


\bibitem[Ye et~al\mbox{.}(2023)]%
        {Ye2023toaster}
\bibfield{author}{\bibinfo{person}{Meryl Ye}, \bibinfo{person}{Eike Schneiders}, \bibinfo{person}{Wen-Ying Lee}, {and} \bibinfo{person}{Malte Jung}.} \bibinfo{year}{2023}\natexlab{}.
\newblock \showarticletitle{The Future of Home Appliances: A Study on the Robotic Toaster as a Domestic Social Robot}. In \bibinfo{booktitle}{\emph{2023 32nd IEEE International Conference on Robot and Human Interactive Communication (RO-MAN)}}. \bibinfo{publisher}{IEEE}, \bibinfo{address}{Busan, Korea}, \bibinfo{pages}{477--482}.
\newblock
\urldef\tempurl%
\url{https://doi.org/10.1109/RO-MAN57019.2023.10309555}
\showDOI{\tempurl}


\bibitem[Yu et~al\mbox{.}(2023)]%
        {Xinyan2023Prediction}
\bibfield{author}{\bibinfo{person}{Xinyan Yu}, \bibinfo{person}{Marius Hoggenm\"{u}ller}, {and} \bibinfo{person}{Martin Tomitsch}.} \bibinfo{year}{2023}\natexlab{}.
\newblock \showarticletitle{Your Way Or My Way: Improving Human-Robot Co-Navigation Through Robot Intent and Pedestrian Prediction Visualisations}. In \bibinfo{booktitle}{\emph{Proceedings of the 2023 ACM/IEEE International Conference on Human-Robot Interaction}} (Stockholm, Sweden) \emph{(\bibinfo{series}{HRI '23})}. \bibinfo{publisher}{Association for Computing Machinery}, \bibinfo{address}{New York, NY, USA}, \bibinfo{pages}{211–221}.
\newblock
\showISBNx{9781450399647}
\urldef\tempurl%
\url{https://doi.org/10.1145/3568162.3576992}
\showDOI{\tempurl}


\bibitem[Yu et~al\mbox{.}(2024)]%
        {Yu2024HelpSeeking}
\bibfield{author}{\bibinfo{person}{Xinyan Yu}, \bibinfo{person}{Marius Hoggenm\"{u}ller}, {and} \bibinfo{person}{Martin Tomitsch}.} \bibinfo{year}{2024}\natexlab{}.
\newblock \showarticletitle{From Agent Autonomy to Casual Collaboration: A Design Investigation on Help-Seeking Urban Robots}. In \bibinfo{booktitle}{\emph{Proceedings of the CHI Conference on Human Factors in Computing Systems}} \emph{(\bibinfo{series}{CHI '24})}. \bibinfo{publisher}{Association for Computing Machinery}, \bibinfo{address}{Honolulu, HI, USA}.
\newblock
\urldef\tempurl%
\url{https://doi.org/10.1145/3613904.3642389}
\showDOI{\tempurl}


\bibitem[Zaga et~al\mbox{.}(2016)]%
        {Zaga2016}
\bibfield{author}{\bibinfo{person}{Cristina Zaga}, \bibinfo{person}{Roelof~A.J. De~Vries}, \bibinfo{person}{Sem~J. Spenkelink}, \bibinfo{person}{Khiet~P. Truong}, {and} \bibinfo{person}{Vanessa Evers}.} \bibinfo{year}{2016}\natexlab{}.
\newblock \showarticletitle{Help-Giving Robot Behaviors in Child-Robot Games: Exploring Semantic Free Utterances}. In \bibinfo{booktitle}{\emph{The Eleventh ACM/IEEE International Conference on Human Robot Interaction}} \emph{(\bibinfo{series}{HRI '16})}. \bibinfo{publisher}{IEEE Press}, \bibinfo{address}{Christchurch, New Zealand}, \bibinfo{pages}{541–542}.
\newblock
\showISBNx{9781467383707}


\bibitem[Zhou and Tian(2020)]%
        {Zhou2020Sad}
\bibfield{author}{\bibinfo{person}{Shujie Zhou} {and} \bibinfo{person}{Leimin Tian}.} \bibinfo{year}{2020}\natexlab{}.
\newblock \showarticletitle{Would you help a sad robot? Influence of robots’ emotional expressions on human-multi-robot collaboration}. In \bibinfo{booktitle}{\emph{2020 29th IEEE International Conference on Robot and Human Interactive Communication (RO-MAN)}}. \bibinfo{publisher}{IEEE}, \bibinfo{address}{Naples, Italy}, \bibinfo{pages}{1243--1250}.
\newblock
\urldef\tempurl%
\url{https://doi.org/10.1109/RO-MAN47096.2020.9223524}
\showDOI{\tempurl}


\bibitem[Zieger et~al\mbox{.}(2023)]%
        {zieger2023happiness}
\bibfield{author}{\bibinfo{person}{Scott Zieger}, \bibinfo{person}{Jiayuan Dong}, \bibinfo{person}{Skye Taylor}, \bibinfo{person}{Caitlyn Sanford}, {and} \bibinfo{person}{Myounghoon Jeon}.} \bibinfo{year}{2023}\natexlab{}.
\newblock \showarticletitle{Happiness and high reliability develop affective trust in in-vehicle agents}.
\newblock \bibinfo{journal}{\emph{Frontiers in Psychology}}  \bibinfo{volume}{14} (\bibinfo{year}{2023}), \bibinfo{pages}{1129294}.
\newblock
\urldef\tempurl%
\url{https://doi.org/10.3389/fpsyg.2023.1129294}
\showDOI{\tempurl}


\bibitem[Zuckerman et~al\mbox{.}(2020)]%
        {Zuckerman2020}
\bibfield{author}{\bibinfo{person}{Oren Zuckerman}, \bibinfo{person}{Dina Walker}, \bibinfo{person}{Andrey Grishko}, \bibinfo{person}{Tal Moran}, \bibinfo{person}{Chen Levy}, \bibinfo{person}{Barak Lisak}, \bibinfo{person}{Iddo~Yehoshua Wald}, {and} \bibinfo{person}{Hadas Erel}.} \bibinfo{year}{2020}\natexlab{}.
\newblock \showarticletitle{Companionship Is Not a Function: The Effect of a Novel Robotic Object on Healthy Older Adults' Feelings of "Being-Seen"}. In \bibinfo{booktitle}{\emph{Proceedings of the 2020 CHI Conference on Human Factors in Computing Systems}} (<conf-loc>, <city>Honolulu</city>, <state>HI</state>, <country>USA</country>, </conf-loc>) \emph{(\bibinfo{series}{CHI '20})}. \bibinfo{publisher}{Association for Computing Machinery}, \bibinfo{address}{New York, NY, USA}, \bibinfo{pages}{1–14}.
\newblock
\showISBNx{9781450367080}
\urldef\tempurl%
\url{https://doi.org/10.1145/3313831.3376411}
\showDOI{\tempurl}


\end{thebibliography}

\appendix

\end{document}